\documentclass[a4paper,11pt]{article}
\pdfoutput=1 

\usepackage{jheppub} 
\usepackage[T1]{fontenc} 
\usepackage{slashed}
\usepackage{arydshln}
\usepackage{caption}
\usepackage{subcaption}
\usepackage{nicematrix}
\usepackage{tikz}
\usepackage{tkz-euclide}
\usepackage[compat=1.1.0]{tikz-feynman}
\usepackage{booktabs}
\usepackage{lipsum}
\usepackage{setspace}
\usepackage{subfiles}
\usepackage{hhline}
\usepackage{stackengine}
\usepackage{multirow}
\usepackage{graphicx} 
\usepackage{array}
\usepackage{makecell}
\usepackage[export]{adjustbox}

\title{\boldmath Revising the Mass of Light Hybrid Mesons: NLO QCD Sum Rules Point to $\phi(2170)$ as a Prime Candidate}

\author[a]{Shuang-Hong Li}
\author[b]{Zhuo-Ran Huang}
\author[c,d]{Wei Chen}
\author[a]{Hong-Ying Jin}

\affiliation[a]{Zhejiang Institute of Modern Physics, School of Physics, Zhejiang University, Hangzhou, 310027, China}
\affiliation[b]{Department of Physics, College of Physics, Mechanical and Electrical Engineering, Jishou University, Jishou 416000, China}
\affiliation[c]{School of Physics, Sun Yat-Sen University, Guangzhou 510275, China}
\affiliation[d]{Southern Center for Nuclear-Science Theory (SCNT), Institute of Modern Physics, Chinese Academy of Sciences, Huizhou 516000, Guangdong Province, China}

\emailAdd{shlee@zju.edu.cn}
\emailAdd{huangzhuoran@126.com}
\emailAdd{chenwei29@mail.sysu.edu.cn}
\emailAdd{jinhongying@zju.edu.cn}

\abstract{We present a comprehensive next-to-leading order (NLO) QCD sum rule analysis for light hybrid mesons with $J^{PC}=1^{--}$, incorporating condensates up to dimension-8 and NLO corrections to the perturbative, gluon condensate, and four-quark condensate contributions. These corrections are found to be substantial and reveal the necessity of contributions beyond leading order. Employing both Laplace (LSR) and Gaussian (GSR) sum rules, our analysis predicts a mass in the conservative range of $2.1-2.4\,\text{GeV}$ for the light $1^{--}$ hybrid. These predictions are significantly lower than previous leading-order (LO) estimates (around $2.9\,\text{GeV}$) and bridge the gap between QCD sum rules and other approaches. Our findings establish the $\phi(2170)$ resonance as a prime candidate for the light vector hybrid meson.}

\begin{document} 	
	\maketitle
	\flushbottom

\section{Introduction}

Hybrid mesons ($q\bar{q}g$), where a gluon acts as a fundamental constituent, are essential for a deeper understanding of Quantum Chromodynamics (QCD) beyond simple $q\bar{q}$ mesons. Although states with exotic quantum numbers often garner the most attention, hybrids with conventional quantum numbers, such as the $J^{PC}=1^{--}$ vector hybrid, also present a compelling and complex puzzle~\cite{Hybrid_mesons,Hybrid_decay,1--_flux_tube,1--_LSR_weyers,1--_GSR_chen,1--_Guo,1--_lattice_dudek,1--_lattice_liu}.

For decades, theoretical predictions for the mass of the $1^{--}$ hybrid have been starkly inconsistent. The flux-tube model suggests masses around $1.8-1.9\,\text{GeV}$ for $u\bar{u}g$ and $2.1-2.2\,\text{GeV}$ for $s\bar{s}g$ ~\cite{1--_flux_tube}, while early QCD sum rule calculations for $s\bar{s}g$ predicted much heavier states near $2.9\,\text{GeV}$~\cite{1--_LSR_weyers}, which was confirmed by a later work~\cite{1--_GSR_chen}. Meanwhile, lattice QCD simulations place the mass in an intermediate range of $2.2-2.3\,\text{GeV}$ for isovector hybrid and $2.4-2.5\,\text{GeV}$ for its isoscalar partner~\cite{1--_lattice_dudek}, and a possible candidate around $2.1-2.3\,\text{GeV}$ noted in Ref~\cite{1--_lattice_liu}.

Experimentally, the $\phi(2170)$ resonance~\cite{pdg,phi_2170_babar_06,phi_2170_babar_08,phi_2170_bes_08,phi_2170_belle_09,phi_2170_bes3_19,barbar_2012} stands out as the most prominent candidate for a strangeonium-like $1^{--}$ hybrid. First observed by BaBar collaboration~\cite{phi_2170_babar_06,phi_2170_babar_08,barbar_2012}, it decays predominantly into final states with a hidden $s\bar{s}$ pair, such as $\phi f_0(980)$, $\phi\pi^+\pi^-$, and $\phi\eta$, suggesting a strangeonium-like nature. However, its measured mass of approximately $2.16\,\text{GeV}$, which aligns well with flux-tube and lattice predictions, is in sharp conflict with the previous, heavier sum rule estimates. This discrepancy has fueled a long-standing controversy over its actual nature, with alternative interpretations including tetraquark states~\cite{1--_tetra_chen_2018,1--_tetra_chen, 1--_tetra_lebed}, $\Lambda\bar{\Lambda}$ molecules~\cite{1--_mole_valery, 1--_mole_li}, a $\phi K\bar{K}$ system or $\phi f_0(980)$ molecule~\cite{1--_phif0_KP, 1--_phif0_mart}, or conventional excited $s\bar{s}$ meson~\cite{1--_ssbar_page, 1--_ssbar_yan}.

To resolve this critical tension, we perform a comprehensive NLO QCD sum rule analysis of the $1^{--}$ hybrid mass. By incorporating previously uncalculated NLO corrections to the perturbative, $\langle GG\rangle$ and $\langle \bar{q}q \rangle^2$ contributions, we present a revised, robust prediction. In the following sections, we detail our NLO calculation and apply both Laplace and Gaussian sum rules to extract the hybrid mass. The resulting masses --- $2.31 \pm 0.23$ GeV for $s\bar{s}g$ and $2.25 \pm 0.23$ GeV for $u\bar{u}g$ --- are substantially lower than previous predictions, making it compatible with the observed $\phi(2170)$.


\section{Formalism for The $1^{--}$ Hybrid Correlator at NLO}\label{qsr_hybrid}

\begin{figure}[h!]
	\centering
	\includegraphics[width=14cm]{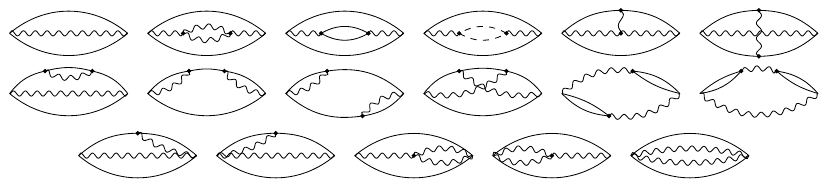}
	\includegraphics[width=13.5cm]{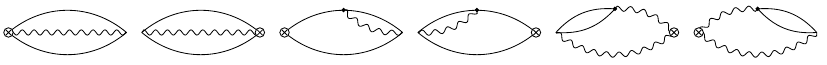}
	\caption{Diagrams of perturbative contributions; the crosses denote the counterterms.}
	\label{d0}
\end{figure}

\begin{figure}[t!]
	\centering
	\includegraphics[width=15cm]{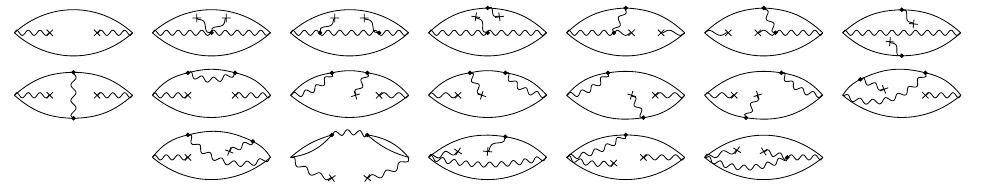}
	\includegraphics[width=13cm]{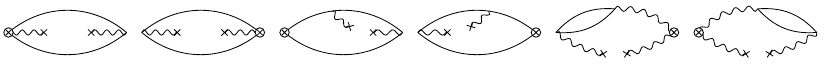}
	\caption{Diagrams of $\langle GG\rangle$ contributions; the crosses denote the counterterms.}
	\label{c_gg}
\end{figure}

\begin{figure}[t!]
	\centering
	\includegraphics[width=15cm]{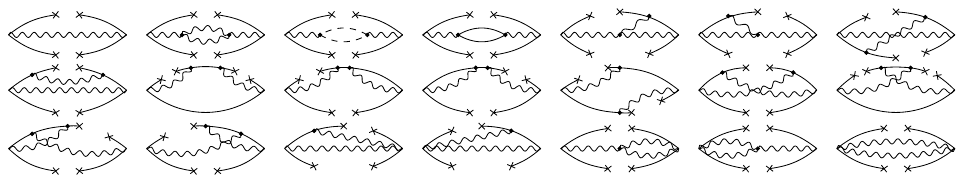}
	\caption{Diagrams of $\langle qq\rangle^2$ contributions.}
	\label{c_qq2}
\end{figure}

We define the interpolating current for the $1^{--}$ hybrid meson as:
\begin{equation}
	J^\mu=g\overline{\Psi} T^n \widetilde{G}^{n\,\rho\mu}\gamma_\rho\gamma^5\Psi.
	\label{J_H_current}
\end{equation}
Here, $\widetilde{G}^{n\,\rho\mu} = \frac{1}{2}\varepsilon^{\rho\mu\alpha\beta}G^n_{\alpha\beta}$, $\varepsilon^{0123}=1$, and $T^n=\lambda^n/2$ is the color matrix. The corresponding $1^{--}$ hybrid mass can be derived from the vector part, $\Pi_v(q^2)$, of the correlator:  
\begin{equation}
	\begin{split}
		\Pi^{\mu\nu}(q^2) &=i\int d^4x\, e^{-iqx}\langle T J^\mu(x)J^{\dagger\,\nu}(0)\rangle\\
		&= \Big(\frac{q^\mu q^\nu}{q^2}-g^{\mu\nu}\Big)\Pi_v(q^2) + \frac{q^\mu q^\nu}{q^2}\Pi_s(q^2).
	\end{split}
\end{equation}

Utilizing operator product expansion (OPE)~\cite{qcd_book,qsr,svz_1,svz_2}, our calculation incorporates condensates up to dimension-8. The perturbative, $\langle GG\rangle$, and $\langle \bar{q}q\rangle^2$ contributions are calculated at next-to-leading order (NLO), as shown in figures~\ref{d0}, \ref{c_gg}, and \ref{c_qq2}. For the diagrams of perturbative and $\langle GG\rangle$ contributions, the hybrid current should be renormalized.

\begin{figure}[h!]
	\centering
	\includegraphics[width=12.5cm]{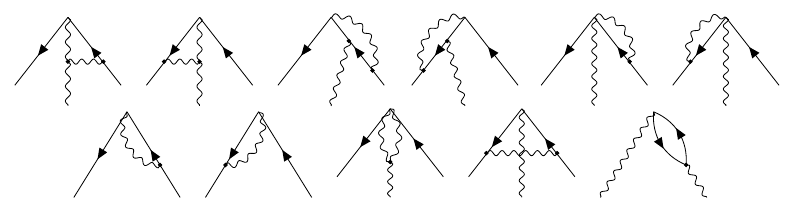}
	\caption{Diagrams involved in hybrid operator renormalization at $O(g^2)$.}
	\label{hybrid_ren}
\end{figure}


\subsection{Renormalization of A Generic Hybrid Current}\label{ren}
Consider a generic hybrid operator
\begin{equation}
	J = g \overline{\Psi}_{f_a}  \Gamma G_{\mu\nu} \Psi_{f_b},
	\label{ap_hybrid}
\end{equation}
where $G_{\mu\nu}= T^n G^n_{\mu\nu}$; $\Gamma$ denotes arbitrary $\gamma$-matrices; $f_a$ and $f_b$ are flavor indices. For the renormalization at $O(g^2)$, it is sufficient to consider the Green's functions:
\begin{equation}
	\begin{split}
		\langle T\,J\,& \Psi_{f_a} A^n_\alpha \overline{\Psi}_{f_b}\rangle,\\
		\langle T\,J\,& \Psi_{f_a}\overline{\Psi}_{f_b}\rangle,\\
		\langle T\,J\,&  A^n_\alpha A^m_\beta\rangle.\\
	\end{split}
	\label{ap_ren_green}
\end{equation}
The corresponding diagrams are shown in figure~\ref{hybrid_ren}. The fermion and gluon fields in Eqs.~\eqref{ap_hybrid} and \eqref{ap_ren_green} are implicitly renormalized; therefore, loops on the external legs need not be considered.

Several observations simplify the determination of the counterterms. The loop integrals in the first, third, fifth, and seventh diagrams of figure~\ref{hybrid_ren} relate only to the $G_{\mu\nu}\Psi_{f_b}$ part of the operator, as:
\begin{subequations}

\begin{equation}\vcenter{\hbox{
	\begin{tikzpicture}[scale=0.75, transform shape]
			\begin{feynman}
				\vertex (a0);
				\vertex [below=1.15cm of a0](a00);
				\vertex [left=0.9cm of a00](aLL);
				\vertex [right=0.9cm of a00](aRR);
				\vertex [below=0.75cm of a0](au);
				\vertex [right=0.6cm of au](aur);
				\vertex [below=1.5cm of a0](auu);
				\diagram*[small]{
					(aRR)--[fermion](a0)--[fermion](aLL),
					(auu)--[photon](au),
					(aur)--[photon](au)--[photon](a0),
				};
				\filldraw(au) circle(0.03cm);
				\filldraw(aur) circle(0.03cm);
			\end{feynman}
		\end{tikzpicture}}}\!+\! \vcenter{\hbox{\begin{tikzpicture}[scale=0.75, transform shape]
			\begin{feynman}
				\vertex (a0);
				\vertex [below=1.15cm of a0](a00);
				\vertex [left=0.9cm of a00](aLL);
				\vertex [right=0.9cm of a00](aRR);
				\vertex [below=0.4cm of a0](au);
				\vertex [right=0.3cm of au](aur);
				\vertex [below=0.9cm of a0](aU);
				\vertex [right=0.7cm of aU](aUr);
				\vertex [below=1.5cm of a0](auu);
				\diagram*[small]{
					(aRR)--[fermion](a0)--[fermion](aLL),
					(auu)--[photon](aur),
					(a0)--[photon, bend left = 70](aUr),
				};
				\filldraw(aur) circle(0.03cm);
				\filldraw(aUr) circle(0.03cm);
			\end{feynman}
		\end{tikzpicture}}}\!+\! \vcenter{\hbox{\begin{tikzpicture}[scale=0.75, transform shape]
			\begin{feynman}
				\vertex (a0);
				\vertex [below=1.15cm of a0](a00);
				\vertex [left=0.9cm of a00](aLL);
				\vertex [right=0.9cm of a00](aRR);
				\vertex [below=0.75cm of a0](au);
				\vertex [right=0.6cm of au](aur);
				\vertex [below=1.5cm of a0](auu);
				\diagram*[small]{
					(aRR)--[fermion](a0)--[fermion](aLL),
					(auu)--[photon](a0),
					(a0)--[photon, bend left = 70](aur),
				};
				\filldraw(aur) circle(0.03cm);
			\end{feynman}
		\end{tikzpicture}}}\!+\!\vcenter{\hbox{\begin{tikzpicture}[scale=0.75, transform shape]
			\begin{feynman}
				\vertex (a0);
				\vertex [below=1.45cm of a0](a00);
				\vertex [left=0.9cm of a00](aLL);
				\vertex [right=0.9cm of a00](aRR);
				\vertex [below=0.4cm of a0](au);
				\vertex [right=0.35cm of au](aur);
				\vertex [below=0.9cm of a0](aU);
				\vertex [right=0.56cm of aU](aUr);
				\vertex [below=1.5cm of a0](auu);
				\diagram*[small]{
					(aRR)--[fermion](a0)--[fermion](aLL),
					(a0)--[photon, bend right = 70](aUr),
				};
				\filldraw(aUr) circle(0.03cm);
			\end{feynman}
	\end{tikzpicture}}} = 
		\vcenter{\hbox{\begin{tikzpicture}[scale=0.75, transform shape]
					\begin{feynman}
						\vertex (a0);
						\vertex [below=1.15cm of a0](a00);
						\vertex [left=0.9cm of a00](aLL);
						\diagram*[small]{
							(a0)--[fermion](aLL)
						};
					\end{feynman}
		\end{tikzpicture}}}\times\Bigg\{\ \
	\vcenter{\hbox{\begin{tikzpicture}[scale=0.75, transform shape]
			\begin{feynman}
				\vertex (a0);
				\vertex [below=1.15cm of a0](a00);
				\vertex [right=0.9cm of a00](aRR);
				\vertex [below=0.75cm of a0](au);
				\vertex [right=0.6cm of au](aur);
				\vertex [below=1.5cm of a0](auu);
				\diagram*[small]{
					(aRR)--[fermion](a0),
					(auu)--[photon](au),
					(aur)--[photon](au)--[photon](a0),
				};
				\filldraw(au) circle(0.03cm);
				\filldraw(aur) circle(0.03cm);
			\end{feynman}
		\end{tikzpicture}}}\!+\! \vcenter{\hbox{\begin{tikzpicture}[scale=0.75, transform shape]
		\begin{feynman}
			\vertex (a0);
			\vertex [below=1.15cm of a0](a00);
			\vertex [right=0.9cm of a00](aRR);
			\vertex [below=0.4cm of a0](au);
			\vertex [right=0.3cm of au](aur);
			\vertex [below=0.9cm of a0](aU);
			\vertex [right=0.7cm of aU](aUr);
			\vertex [below=1.5cm of a0](auu);
			\diagram*[small]{
				(aRR)--[fermion](a0),
				(auu)--[photon](aur),
				(a0)--[photon, bend left = 70](aUr),
			};
			\filldraw(aur) circle(0.03cm);
			\filldraw(aUr) circle(0.03cm);
		\end{feynman}
		\end{tikzpicture}}}+ \vcenter{\hbox{\begin{tikzpicture}[scale=0.75, transform shape]
		\begin{feynman}
			\vertex (a0);
			\vertex [below=1.15cm of a0](a00);
			\vertex [right=0.9cm of a00](aRR);
			\vertex [below=0.75cm of a0](au);
			\vertex [right=0.6cm of au](aur);
			\vertex [below=1.5cm of a0](auu);
			\diagram*[small]{
				(aRR)--[fermion](a0),
				(auu)--[photon](a0),
				(a0)--[photon, bend left = 70](aur),
			};
			\filldraw(aur) circle(0.03cm);
		\end{feynman}
		\end{tikzpicture}}}\!+\!\vcenter{\hbox{\begin{tikzpicture}[scale=0.75, transform shape]
		\begin{feynman}
			\vertex (a0);
			\vertex [below=1.45cm of a0](a00);
			\vertex [right=0.9cm of a00](aRR);
			\vertex [below=0.4cm of a0](au);
			\vertex [right=0.35cm of au](aur);
			\vertex [below=0.9cm of a0](aU);
			\vertex [right=0.56cm of aU](aUr);
			\vertex [below=1.5cm of a0](auu);
			\diagram*[small]{
				(aRR)--[fermion](a0),
				(a0)--[photon, bend right = 70](aUr),
			};
			\filldraw(aUr) circle(0.03cm);
		\end{feynman}
	\end{tikzpicture}}}\ \Bigg\}.
\end{equation}

	Therefore, the corresponding counterterm can be written as:
	\begin{equation}
		\frac{1}{\varepsilon}\overline{\Psi}\,\Gamma \times \{\text{operators acting on }\Psi\}.
	\end{equation}
\end{subequations}

Similarly, the loop integral in the ninth diagram in figure~\ref{hybrid_ren} is related only to $G^n_{\mu\nu}$, whereas the loop integrals in the last two diagrams in figure~\ref{hybrid_ren} are irrelevant to $G^n_{\mu\nu}$. In each case, the number of diagrams involved is much smaller than the total number of diagrams shown in figure~\ref{hybrid_ren}. Furthermore, this allows us to construct a renormalized hybrid operator for arbitrary $\gamma$-matrices. The result is:
\begin{equation}
	\begin{split}
		J_r =&\,\,(1 + C_0 + C_4) Z_2^{-1} Z_3^{-\frac{1}{2}}\,\overline{\Psi}_{f_a} \,g \Gamma G^{\mu\nu} \Psi_{f_b} \\
		&+ C_1\,\Big(\overline{\Psi}_{f_a}\big[\big(\overleftarrow{\nabla}^\mu\overleftarrow{\nabla}^\rho + \overleftarrow{\nabla}^\rho\overleftarrow{\nabla}^\mu\big){\sigma_\rho}^\nu.\Gamma + \Gamma.{\sigma^\nu}_\rho\big(\nabla^\mu\nabla^\rho + \nabla^\rho \nabla^\mu\big)\big]\Psi_{f_b} -\{\mu\leftrightarrow\nu\}\Big)\\
		&+ C_2\,\Big(\overline{\Psi}_{f_a}\big[\big(\overleftarrow{\slashed{\nabla}}-i\, m_{f_a}\big)\gamma^\nu.\Gamma A^\mu  g + g A^\mu \Gamma.\gamma^\nu.\big(\slashed{\nabla}+i\, m_{f_b}\big)\big]\Psi_{f_b}-\{\mu\leftrightarrow\nu\}\Big)\\
		&+ C_3\,\overline{\Psi}_{f_a}\big(\overleftarrow{\nabla}^2\sigma^{\mu\nu}.\Gamma + \Gamma.\sigma^{\mu\nu}\nabla^2\big)\Psi_{f_b}\\
		&+ C_5\, \overline{\Psi}_{f_a} \big(\gamma^5.\Gamma + \Gamma.\gamma^5\big)i\,\varepsilon^{\mu\nu\rho\sigma}G_{\rho\sigma}\Psi_{f_b}\\
		&+ C_6\,\Big(\overline{\Psi}_{f_a}\big[i\sigma^{\rho\mu}.\Gamma + i \Gamma.\sigma^{\mu\rho}\big]{G_\rho}^\nu\Psi_{f_b} -\{\mu\leftrightarrow\nu\}\Big)\\
		&+ C_7\,\Big(\overline{\Psi}_{f_a}\big[\overleftarrow{\nabla}^\mu \gamma^\nu.\Gamma m_{f_a} + m_{f_b}\Gamma.\gamma^\nu \nabla^\mu\big]\Psi_{f_b}-\{\mu\leftrightarrow\nu\}\Big)\\
		&+ C_8\, \overline{\Psi}_{f_a}\big(\sigma^{\mu\nu}.\Gamma m_{f_a}^2 + m_{f_b}^2 \Gamma.\sigma^{\mu\nu}\big)\Psi_{f_b}\\
		&+ C_9\, \overline{\Psi}_{f_a}\gamma^\rho.\gamma^\sigma.\Gamma.\gamma_\sigma.\gamma_\rho G_{\mu\nu}\Psi_{f_b}\\
		&+ C_{10}\,\delta_{f_af_b}\Big(\text{Tr}\big[\Gamma.\sigma^{\alpha\beta}\big]m_{f_a} \,G^{n\,\mu\nu}G^n_{\alpha\beta}+\frac{2}{3}\text{Tr}\big[\Gamma.\gamma_\alpha\big]G^{n\,\mu\nu}D^{nm}_\beta G^{m\,\alpha\beta}\Big).
	\end{split}
	\label{ren_hybrid}
\end{equation}
Here, the fields on the right-hand side are bare fields. The
\begin{equation}
	Z_2 = 1-\frac{g^2 C_F}{16\pi^2\varepsilon}\quad\text{and}\quad Z_3=1-\frac{g^2}{16\pi^2\varepsilon}\Big[-\frac{5}{3}C_A+\frac{2}{3}n_f\Big]
	\label{Z_constant}
\end{equation}
are fermion and gluon renormalization constants (with $d=4-2\varepsilon$) in Feynman gauge. The values of $C_n$ in Feynman gauge are listed in table~\ref{hybrid_ren_c}. Massive quark propagators are used to obtain these values; the term $\propto C_8$ vanishes in the $m^2\rightarrow0$ limit.

The term $\propto C_2$ is not gauge invariant but vanishes when applying the equations of motion, as required for non-gauge-invariant counterterms~\cite{book_ren}. The $C_0$ term corresponds to the third-to-last diagram in figure~\ref{hybrid_ren}, while the $C_9$ and $C_{10}$ terms correspond to the last two diagrams. We do not combine $C_0$ and $C_4$ into a single constant because the $C_0$ term is directly related to the cancellation of the IR pole, as discussed in Appendix~\ref{gluon_IR_pole}. For the $\langle GG\rangle$ diagrams shown in figure~\ref{c_gg}, the renormalized coupling constant $g$ should be written explicitly:
\begin{equation}
	g = Z_g^{-1}g_0=\big[1+\frac{g^2}{16\pi^2\varepsilon}\big(\frac{11}{6}C_A-\frac{1}{3}n_f\big)\big]g_0,
	\label{g_Z}
\end{equation}
which ensures the cancellation of the $\log/\varepsilon$ pole and the $n_f$-dependence in $Z_3$, since no $n_f$-dependent diagram is involved in the $\langle GG\rangle$ diagrams.

\begin{table}[h!]
	\centering
	\caption{The coefficients of each counterterm in eq.~\eqref{ren_hybrid}, obtained under Feynman gauge.\label{hybrid_ren_c}}
	\renewcommand{\arraystretch}{1}
	\stackengine{0pt}{
		\begin{tabular}{|w{c}{1.4cm}|w{c}{1.4cm}|w{c}{1.4cm}|w{c}{1.4cm}|w{c}{1.4cm}|w{c}{1.4cm}|}
			\noalign{\hrule height 1pt}
			$C_0$&$C_1$&$C_2$&$C_3$&$C_4$&$C_5$\\
			\noalign{\hrule height 0.2pt}
			\Gape[2pt][2pt]{$\frac{3C_Ag}{32\pi^2\varepsilon}$}&$\frac{C_F g^2}{192\pi^2\varepsilon}$&$\frac{C_A g^2}{128\pi^2 \varepsilon}$&$\frac{C_F g^2}{96\pi^2\varepsilon}$&$-\frac{C_Ag}{16\pi^2\varepsilon}$&$\frac{C_Ag^2}{128\pi^2\varepsilon}$\\
			\noalign{\hrule height 1pt}
		\end{tabular}
	}{
		\begin{tabular}{|w{c}{1.4cm}|w{c}{1.4cm}|w{c}{1.4cm}|w{c}{1.4cm}|w{c}{1.4cm}|}
			$C_6$&$C_7$&$C_8$&$C_9$&$C_{10}$\\ 
			\noalign{\hrule height 0.2pt}
			\Gape[2pt][2pt]{$\frac{(3C_A^2 +1)g^2}{128C_A \pi^2\varepsilon}$}&$\frac{C_F g^2}{32\pi^2\varepsilon}$&$\frac{C_F g^2}{32\pi^2\varepsilon}$&$\frac{g^2}{128C_A\pi^2\varepsilon}$&$\frac{g^2}{64\pi^2\varepsilon}$\\
			\noalign{\hrule height 1pt}
		\end{tabular}
	}{U}{l}{F}{F}{S}
	\renewcommand{\arraystretch}{1}
\end{table}


\subsection{The $1^{--}$ Hybrid Correlator at NLO \label{NLO_correlator_1--}}

\vspace{-0.2cm}
\begin{figure}[h!]
	\centering
	\begin{subfigure}{0.25\textwidth}
		\centering
		\includegraphics[width=2.2cm]{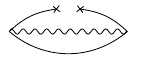}
		\vspace{-0.2cm}
		\caption{$m\langle \bar{q}q\rangle$.}
	\end{subfigure}
	\hfill
	\begin{subfigure}{0.65\textwidth}
		\centering
		\includegraphics[width=6.6cm]{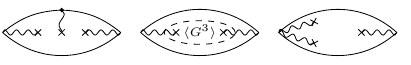}
		\vspace{-0.2cm}
		\caption{$\langle GGG\rangle$.}
	\end{subfigure}
	\vspace{-0.2cm}
	\caption{Diagrams corresponding to $m\langle \bar{q}q\rangle$ and $\langle GGG\rangle$ contributions; eq.\eqref{con_d6_2g} is used for $\langle GGG\rangle$ contribution. }
	\label{c_mqq_ggg}
\end{figure}

\vspace{-0.3cm}

\begin{figure}[h!]
	\centering
	\includegraphics[width=9.6cm]{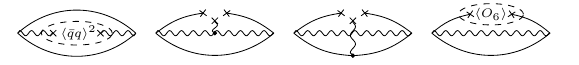}
	\vspace{-0.2cm}
	\caption{Diagrams corresponding to $\langle qq\rangle^2$ and $m\langle qGq\rangle$ contributions. Terms such as $\langle \bar{q}^a_iG_{\alpha\beta}\nabla_\nu q^b_j\rangle$ and $\langle \bar{q}^a_iD_\alpha G^n_{\beta\mu} q^b_j\rangle$ can be expressed in terms of $\langle qq\rangle^2$ and $m\langle qGq\rangle$ by eq.~\eqref{con_d6_2q}.}
	\label{c_qq2mqgq}
\end{figure}

\vspace{-0.3cm}

\begin{figure}[h!]
	\centering
	\includegraphics[width=12cm]{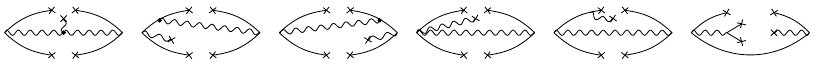}
	\vspace{-0.2cm}
	\caption{Diagrams corresponding to $\langle \bar{q}q\rangle\langle \bar{q}Gq\rangle$ contribution.}
	\label{c_d8}
\end{figure}

The diagrams involved in the evaluation of $1^{--}$ Hybrid Correlator are shown in figures~\ref{d0}-\ref{c_qq2} and \ref{c_mqq_ggg}-\ref{c_d8}. The $\langle GGG\rangle$ contribution does not contribute to the imaginary part of the correlator. The $\gamma^5$ is treated in this way: for loops involving an odd number of $\gamma^5$, the BMHV scheme~\cite{BMHV1,feyncalc} is adopt, while for loops involving an even number of $\gamma^5$, we simply anticommute them (NDR scheme~\cite{feyncalc}). The cancellation of infrared divergences, and the evaluation of $\langle\bar{q}q\rangle\langle\bar{q}Gq\rangle$ are discussed in Appendix~\ref{gluon_IR_pole} and \ref{d8_factor}, respectively.

To obtain the $\langle qq\rangle^2$ and $m\langle qGq\rangle$ contributions from figure~\ref{c_qq2mqgq}, the following identities are used~\cite{high_order_condensates}:
\vspace{-0.2cm}
\begin{equation}
	\begin{split}
		\langle \bar{q}_i^a \nabla_\alpha \nabla_\beta\nabla_\nu q_j^b\rangle =\frac{i\delta^{ab}}{C_A}\times\Big[&\frac{C_F\,g^2}{32C_A(d-1)}\langle\bar{q}q\rangle^2\Big(\frac{-g_{\alpha\beta}\gamma_\nu+(d+1)g_{\alpha\nu}\gamma_\beta-g_{\beta\nu}\gamma_\alpha}{d+2}-\frac{\gamma_{\alpha\beta\nu}}{d-2}\Big)\\
		&-m\langle \bar{q}Gq\rangle\Big(\frac{g_{\alpha\beta}\gamma_\nu+g_{\alpha\nu}\gamma_\beta+g_{\beta\nu}\gamma_\alpha}{8d(d+2)}+\frac{\gamma_{\alpha\beta\nu}}{8d(d-1)(d-2)}\Big)\\
		&+m^3\langle\bar{q}q\rangle \frac{\gamma_\alpha g_{\beta\nu}+\gamma_\beta g_{\alpha\nu} + \gamma_\nu g_{\alpha\beta}}{4d(d+2)}\Big]_{ji}\,\, ,
	\end{split}
	\label{con_d6_2q}
\end{equation}
\vspace{-0.2cm}
\begin{equation}
	\begin{split}
		\langle g^2 G^n_{\mu\nu} D_\alpha D_\beta G^m_{\rho\sigma}\rangle =&\frac{\delta^{nm}}{C_A C_F d(d-1)(d^2-4)}\times\\
		&\Big[-(d+2)\langle O^6_1\rangle
		(\!(\!(g_{\sigma\mu}g_{\nu\alpha}g_{\beta\rho}\!-\!(\mu\leftrightarrow\nu))\!-\!(\rho\leftrightarrow\sigma))\!-\!(\alpha\leftrightarrow\beta))\\
		&\quad+\Big[(d-4)\langle O^6_1\rangle+(d-2)\langle O^6_2\rangle\Big]\\
		&\qquad\times\big(((g_{\sigma\mu}g_{\nu\alpha}g_{\beta\rho}-(\mu\leftrightarrow\nu))-(\rho\leftrightarrow\sigma))+(\alpha\leftrightarrow\beta)\big)\\
		&\quad-\big[4(d-1)\langle O^6_1\rangle+2(d-2)\langle O^6_2\rangle\big]
		g_{\alpha\beta}(g_{\mu\rho}g_{\nu\sigma}-(\rho\leftrightarrow\sigma)) \Big],
	\end{split}
	\label{con_d6_2g}
\end{equation}
where
\begin{equation}
	\begin{split}
		\langle O^6_1\rangle =& -\frac{1}{4}\langle g^3f^{abc}G^a_{\eta\mu}G^b_{\mu\nu}G^c_{\nu\eta}\rangle,\\
		\qquad\langle O^6_2\rangle=&-\frac{dC_F}{8C_A}g^2\sum_q\langle \bar{q}q\rangle^2 ,
	\end{split}
\end{equation}
the $\sum_q$ denotes the sum over the flavors, and the vacuum saturation hypothesis~\cite{qcd_book,vs_hypothesis} is adopted to write the dimension-6 quark condensate as $\langle \bar{q}q\rangle^2$.

Utilizing eq.~\eqref{ap_ren_green}, the $1^{--}$ hybrid correlator at NLO yields:
\begin{align}
		\Pi_v(q^2)=&-\log\!
		\Big(\!\!-\!\frac{q^2}{\mu ^2}\Big)\bigg[\frac{(6 n_f-35) \alpha _s^2 q^6 }{17280 \pi ^4} \log\!\Big(\!\!-\!\frac{q^2}{\mu ^2}\Big)-\frac{(454 n_f-6815) \alpha _s^2 q^6 }{172800 \pi ^4}+\frac{\alpha_s q^6 }{240\pi^3}\bigg]\nonumber\\
		&+\!\bigg[\frac{2 \alpha _s q^2 }{81 \pi ^2}\log\!
		\Big(\!\!-\!\frac{q^2}{\mu ^2}\!\Big)\!-\!\frac{17 \alpha_s q^2 }{864 \pi ^2}\!+\!\frac{q^2}{36 \pi }\bigg]\! \log\!
		\Big(\!\!-\!\frac{q^2}{\mu ^2}\!\Big) \langle GG\rangle \!-\!\frac{4 \alpha _s }{9 \pi }q^2  \log\!
		\Big(\!\!-\!\frac{q^2}{\mu ^2}\!\Big) m\!\left\langle \overline{\Psi}\Psi\text{}\right\rangle\nonumber\\
		&+\frac{1}{162} (24 n_f\!+\!103) \alpha _s^2 \log\! \Big(\!\!-\!\frac{q^2}{\mu
			^2}\Big)\! \left\langle \overline{\Psi}\Psi\text{}\right\rangle ^2\!+\!\frac{15}{32 \pi}\alpha _s \log\!
			\Big(\!\!-\!\frac{q^2}{\mu ^2}\!\Big) m\!\left\langle \overline{\Psi}G\Psi\text{}\right\rangle\!	\label{Pi_v(s)}\\
		&\!-\!\frac{4}{27} \alpha _s^2 \log\!
		\Big(\!\!-\!\frac{q^2}{\mu ^2}\Big) \big[\!\left\langle \bar{u}u\text{}\right\rangle ^2\!\!+\!\left\langle 
		\bar{d}d\text{}\right\rangle ^2\!\!+\!\left\langle 
		\bar{s}s\text{}\right\rangle ^2\!\big]\!-\!\frac{131 \pi  \alpha _s }{162 q^2}\left\langle \overline{\Psi}\Psi\text{}\right\rangle  \left\langle \overline{\Psi}G\Psi\text{}\right\rangle . \nonumber
\end{align}

\vspace{-0.2cm}
\begin{align}
		\Pi_s(q^2)=&-\log\!
		\Big(\!\!-\!\frac{q^2}{\mu ^2}\Big)\bigg[\frac{ \left(6 n_f-35\right)\alpha _s^2q^6 }{34560 \pi ^4}\log\! \Big(\!\!-\!\frac{q^2}{\mu ^2}\Big)-\frac{\left(1722 n_f-20945\right)\alpha _s^2 q^6 }{1036800 \pi ^4}+\frac{\alpha _sq^6 }{480 \pi ^3}\bigg]\nonumber\\
		&-\!\bigg[\frac{\alpha_s q^2 }{27 \pi ^2}\log\!
		\Big(\!\!-\!\frac{q^2}{\mu ^2}\Big)-\frac{115 \alpha_s q^2 }{1728 \pi ^2}+\frac{q^2 }{24 \pi }\bigg]\log\! \Big(\!\!-\!\frac{q^2}{\mu ^2}\Big)\langle GG\, \rangle -\!\frac{\alpha_s }{3 \pi } q^2\log\!
		\Big(\!\!-\!\frac{q^2}{\mu ^2}\Big)m\!\left\langle\overline{\Psi}\Psi\text{}\right\rangle
		\nonumber\\
		& +\frac{37}{162} \alpha_s^2 \log\! \Big(\!\!-\!\frac{q^2}{\mu ^2}\Big) \left\langle\overline{\Psi}\Psi\text{}\right\rangle
		^2+\frac{4}{27} \alpha_s^2 \log\!
		\Big(\!\!-\!\frac{q^2}{\mu ^2}\Big)\big[ \left\langle\bar{u}u\text{}\right\rangle ^2 + \left\langle\bar{d}d\text{}\right\rangle ^2+ \left\langle\bar{s}s\text{}\right\rangle ^2\big]	\label{Pi_s(s)}\\
		&-\frac{15}{32 \pi } \alpha _s \log\! \Big(\!\!-\!\frac{q^2}{\mu ^2}\Big) m\!
		\left\langle\overline{\Psi}G\Psi\text{}\right\rangle-\frac{2 \pi \alpha _s}{27 q^2} \left\langle\overline{\Psi}\Psi\text{}\right\rangle \left\langle\overline{\Psi}G\Psi\text{}\right\rangle . \nonumber
\end{align}

Here, terms that do not contribute to the imaginary part have been omitted, and we adopt $n_f=3$. The $\langle\bar{u}u\text{}\rangle^2+\langle\bar{d}d\text{}\rangle^2+\langle\bar{s}s\text{}\rangle^2$ term originates from the gluon equation of motion, and
\begin{equation*}
	\langle GG\rangle = \langle\alpha_s G^n_{\mu\nu}G^{n\,\mu\nu}\rangle,\quad \langle \overline{\Psi} G \Psi\rangle = \langle \overline{\Psi}\sigma^{\mu\nu}T^n g G^n_{\mu\nu} \Psi\rangle,
\end{equation*}
where $\sigma^{\mu\nu}=\frac{i}{2}[\gamma^\mu,\gamma^\nu]$.

Through the dispersion relation~\cite{qcd_book,qsr,qsr_intro_2,qsr_book}, and under the ``pole $+$ continuum'' ansatz, we have:
\begin{equation}
	\frac{1}{\pi}\text{Im}\Pi(s) = f^2\delta(s-m^2)+\theta(s-s_0)\rho(s),
	\label{pole_continue}
\end{equation}
where $m$ is the mass of the lowest-lying resonance, $s_0$ is the continuum threshold, and $\rho(s)$ is the continuum spectral density.

For the numerical analysis, we use the following quark masses~\cite{pdg} and condensate values~\cite{qcd_sum_review} (both at $\mu=2\,\text{GeV}$):
\begin{align*}
		m_u &= 2.16\pm0.07\text{MeV},& m_d &= 4.70\pm0.07\text{MeV},& m_s &= 93.5\pm 0.8\text{MeV},\\ 
		\langle GG\rangle & = 0.07\pm0.02\,\text{GeV}^4,& \langle \bar{q}q\rangle &= -(0.276)^3\,\text{GeV}^3,& \langle\bar{s}s\rangle &= 0.74\langle \bar{q}q\rangle,\\ 
		\langle\bar{q}Gq\rangle &= M_0^2\langle \bar{q}q\rangle,& \langle\bar{s}Gs\rangle &= M_0^2\langle \bar{s}s\rangle,& M_0^2&=0.8\pm0.2\,\text{GeV}^2,\\
	\label{qcd_parameters}
\end{align*}
where $q=u,d$. The one-loop approximation of running $\alpha_s$ is used:
\begin{equation}
	\alpha_s(\mu^2)=\frac{\alpha_s(m^2_\tau)}{1+\frac{\beta_0}{4\pi}\alpha_s(m^2_\tau)\log\big(\frac{\mu^2}{m^2_\tau}\big)},
\end{equation}
where~\cite{pdg}
\begin{equation*}
	m_\tau=1776.93\pm0.09\text{MeV},\quad \alpha_s(m^2_\tau)=0.314\pm0.014,
	\label{alpha_s}
\end{equation*}
and $\beta_0=9$ for $n_f=3$.


\subsection{Vector Meson---Hybrid Correlator}\label{rho_H}

\begin{figure}[t!]
	\centering
	\includegraphics[width=15.5cm]{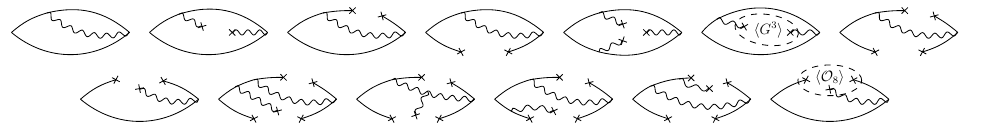}
	\vspace{-0.2cm}
	\caption{Diagrams involved in the calculation of $\langle J^\mu_\rho (x) J^{\dagger\,\nu}(0) \rangle$; eq.~\eqref{d8_expansion} is used for the last diagram.}
	\label{rho-H_diagrams}
\end{figure}

In eq.~\eqref{pole_continue}, we assume only one resonance is involved; however, the vector meson like $\rho$ or $\phi$ also has quantum number $J^{PC}=1^{--}$ and could couple to hybrid current~\eqref{J_H_current}. To investigate the influence of vector meson, the eq.~\eqref{pole_continue} should be modified into a two-resonance spectrum:
\begin{equation}
	\frac{1}{\pi}\text{Im}\Pi(s) = f_1^2\,\delta(s-m_1^2)+f_2^2\delta(s-m_2^2)+\theta(s-s_0)\rho(s).
	\tag{2.14'}
	\label{2pole_continue}
\end{equation} 
Here, $m_1$ and $m_2$ are masses of vector meson and hybrid, respectively; while $f_1$ and $f_2$ represent the coupling strengths of $J^\mu$ to the vector meson and the hybrid, respectively, defined as:
\begin{equation}
	\langle 0| J^\mu|V(q)\rangle=f_1\epsilon^\mu(q)\qquad\text{and}\qquad\langle 0| J^\mu|H(q)\rangle=f_2 \epsilon^\mu(q),
	\label{H_coupling}
\end{equation}
where $V(q)$ and $H(q)$ are vector meson and hybrid states. The $f_1$ can be estimated from the off-diagonal correlator~\cite{rho_1--}:
\begin{equation}
	\begin{split}
		\Pi^{\mu\nu}(q^2) &=i\int d^4x\, e^{-iqx}\langle T J_V^\mu(x)J^{\dagger\,\nu}(0)\rangle\\
		&= \Big(\frac{q^\mu q^\nu}{q^2}-g^{\mu\nu}\Big)\Pi_v^{V H}(q^2) + \frac{q^\mu q^\nu}{q^2}\Pi_s^{V H}(q^2),
		\label{rho_H_correlator}
	\end{split}
\end{equation}
where $J^\mu_V = \overline{\Psi}\gamma^\mu \Psi$ is the vector meson current. The diagrams involved at leading order are shown in figure.~\ref{rho-H_diagrams}, where condensates up to dimension-8 are included. The vector part of eq.~\eqref{rho_H_correlator} yields 
\begin{align}
		\Pi_v^{V H}(q^2) =&-\frac{q^4  }{18 \pi ^3}\alpha_s\log\! \Big(\!\!-\!\frac{q^2}{\mu ^2}\Big)-\frac{4}{9 \pi }\alpha_s\log\! \Big(\!\!-\!\frac{q^2}{\mu ^2}\Big) m \langle\overline{\Psi}\Psi\rangle -\frac{2}{3 q^2} m
		\langle\overline{\Psi}\text{G}\Psi\rangle +\frac{32 \pi }{27 q^2} \alpha_s \langle\overline{\Psi}\Psi\rangle ^2\nonumber\\
		&+\frac{1}{96 \pi ^2 q^2}\log\! \Big(\!\!-\!\frac{q^2}{\mu ^2}\Big) \langle G^3\rangle -\frac{7 }{192 \pi ^2 q^2}\langle G^3\rangle -\frac{185 \pi }{486 q^4} \alpha_s \langle\overline{\Psi}\Psi\rangle  \langle\overline{\Psi}\text{G}\Psi\rangle.
		\label{rho-Hv_correlator}
\end{align}

\section{QCD Sum Rules Analysis}

\subsection{Laplace Sum Rules\label{LSR}}

The Laplace sum rule (LSR) involves applying the Borel transformation~\cite{qcd_book,qsr,qsr_intro_2} to the correlation function. This procedure eliminates subtraction terms from the dispersion relation and exponentially suppresses contributions from higher states and the continuum, thereby isolating the contribution of the lowest-lying resonance. Applying the Borel transformation to the moments of the spectral functions with single resonance yields:

\begin{equation}
	\mathcal{M}^n(\tau,s_0)=\frac{1}{\pi}\int_0^{s_0}ds\ s^n e^{-\tau s}\, \text{Im}\Pi_v(s)=f^2 m^{2n}e^{-\tau m^2},
	\label{moment}
\end{equation}
where $\tau$ is the Borel parameter; the renormalization-group-improved moments are obtained by setting $\mu^2\rightarrow 1/\tau$~\cite{qsr_laplace}. The mass of the lowest resonance can then be extracted from the ratio of these moments:
\begin{equation}
	\mathcal{R}^n(\tau,s_0)=\frac{\mathcal{M}^{n+1}(\tau,s_0)}{\mathcal{M}^n(\tau,s_0)}=m^2.
	\label{ratio}
\end{equation}
In practice, the stability criterion is commonly adopted for determining the mass, which involves identifying the extremum or the most stable region in the curve of $\sqrt{\mathcal{R}^n}(\tau)$.

For a spectrum involving two resonances as in eq.~\eqref{2pole_continue}, to extract the mass of the hybrid, the moment~\eqref{moment} should be modified. Consider
\begin{equation}
	\mathcal{M}^0(\tau,s_0)=\frac{1}{\pi}\int_0^{s_0}ds\ s^n e^{-\tau s}\, \text{Im}\Pi_v(s)=f_1^2\, e^{-\tau m_1^2}+f_2^2 e^{-\tau m_2^2},
	\tag{3.1'}
	\label{moment_rho}
\end{equation} 
instead of introducing coupling constants $f_1$ and $f_2$ explicitly, one can define a modified moment as: 
\begin{equation}
	\widetilde{\mathcal{M}}(\tau,s_0)\equiv\mathcal{M}^0(\tau,s_0)e^{m_1^2 \tau}=f_1^2+f_2^2 e^{(m_1^2- m_2^2)\tau}.
	\label{moment_rho_m}
\end{equation} 
Therefore, the mass information can be derived from the ratio:
\begin{equation}
	\widetilde{\mathcal{R}}(\tau,s_0)= \frac{\partial^2_\tau \widetilde{\mathcal{M}}(\tau,s_0)}{\partial_\tau \widetilde{\mathcal{M}}(\tau,s_0)}=m_1^2-m_2^2.
	\label{ratio_rho}
\end{equation}
Since the mass of the vector meson $m_1$ is well-established, the hybrid mass $m_2$ can be extracted directly from eq.~\eqref{ratio_rho}.

\begin{figure}[t!]
	\centering
	\includegraphics[width=7.2cm,height=4.02cm]{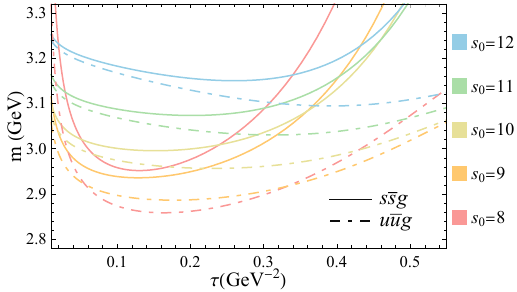}
	\hspace{\fill}
	\hspace{0.2cm}\includegraphics[width=7.2cm]{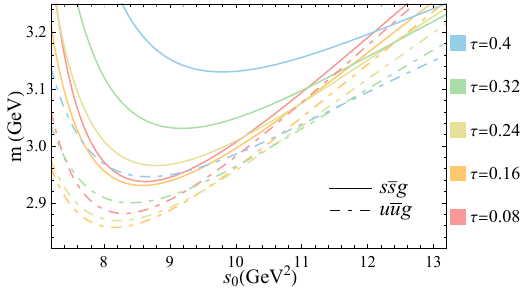}
	\vspace*{-0.2cm}
	\caption{LO single-resonance LSR : $\sqrt{\mathcal{R}^0}(\text{GeV})$ versus $\tau(\text{GeV}^{-2})$ and $s_0(\text{GeV}^2)$.}
	\label{LSR_r0_d8_LO}
\end{figure}

\begin{figure}[t!]
	\centering
	\includegraphics[width=7.2cm]{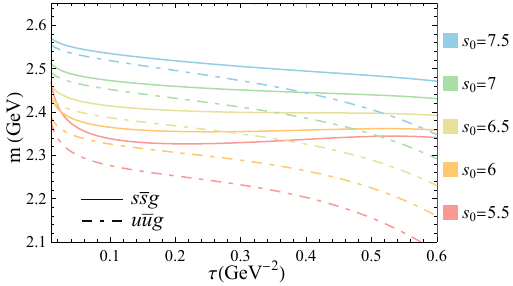}
	\hspace{\fill}
	\includegraphics[width=7.2cm,trim={-0.05cm 0 0 0}]{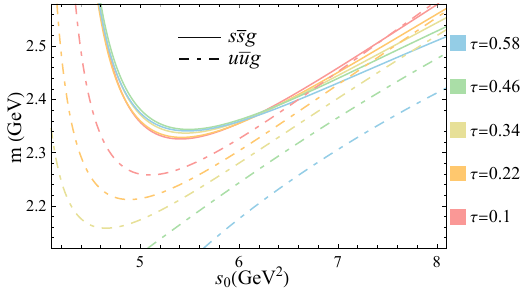}
	\vspace*{-0.2cm}
	\caption{NLO single-resonance LSR: $\sqrt{\mathcal{R}^0}(\text{GeV})$ versus $\tau(\text{GeV}^{-2})$ and $s_0(\text{GeV}^2)$.}
	\label{LSR_r0_d8}
\end{figure}

At leading-order (LO), the single resonance analysis reproduces previous results~\cite{1--_LSR_weyers, 1--_GSR_chen}, predicting a mass $\sim 3\,\text{GeV}$, as shown in figure~\ref{LSR_r0_d8_LO}. However, the inclusion of next-to-leading-order (NLO) corrections substantially lowers this prediction. The NLO single resonance estimation, presented in figure~\ref{LSR_r0_d8}, shows that $s\bar{s}g$ has a mass of $\simeq 2.35\,\text{GeV}$. For $u\bar{u}g$, the $\sqrt{\mathcal{R}^0}(\tau)$ is nearly stable around $\tau\sim0.2\,{\small\text{GeV}^{-2}}$ when $s_0\sim 6\,{\small\text{GeV}^2}$, suggesting a mass of $\sim2.3\,\text{GeV}$.

\begin{figure}[t!]
	\centering
	\includegraphics[width=7.05cm]{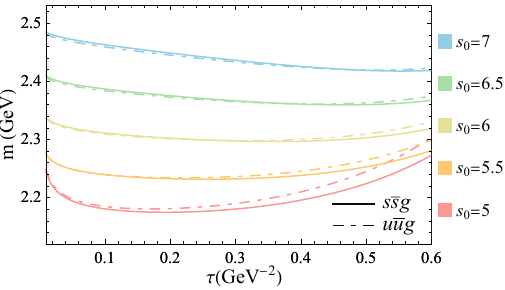}
	\hspace{\fill}
	\includegraphics[width=7.05cm]{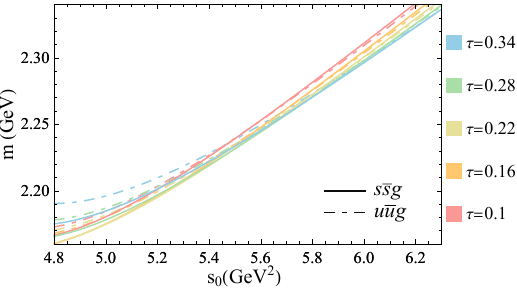}
	\vspace{-0.2cm}
	\caption{NLO LSR: hybrid mass verses $\tau({\small\text{GeV}^{-2}})$ and ${\small s_0(\text{GeV}^2)}$ based on eq.~\eqref{ratio_rho}.}
	\label{rho-ratio_k11}
\end{figure}

Similarly, by setting $m_1$ to be the mass of the $\rho$ or $\phi$ meson, the modified moment (eq.~\eqref{ratio_rho}) yields $m_2\sim2.3\text{GeV}$ for both $\bar{u}ug$ and $\bar{s}sg$, as shown in figure~\ref{rho-ratio_k11}. Compared to the basic LSR, the modified LSR yields more stable results. The consistency between the two NLO results implies that the vector meson's influence is small.

\begin{figure}[t!]
	\raggedright
	\begin{subfigure}{0.48\textwidth}
		\includegraphics[width=7.2cm]{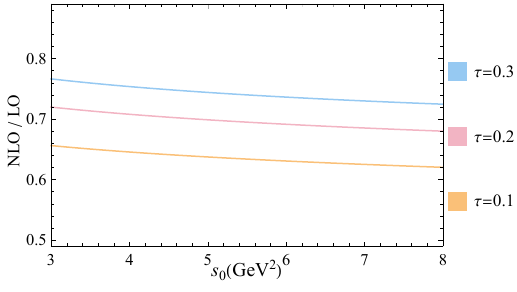}
		\caption{Perturbative terms.}
	\end{subfigure}
	\hspace*{\fill}	
	\begin{subfigure}{0.48\textwidth}
		\includegraphics[width=7.2cm]{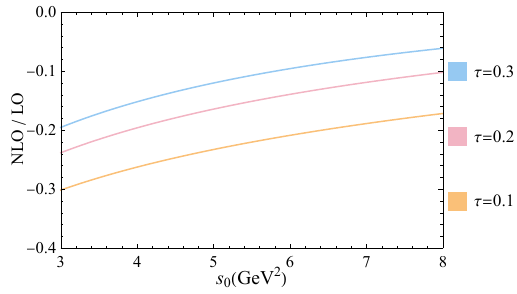}
		\caption{$\langle GG\rangle$ terms.}
	\end{subfigure}\\
	\caption{Ratio of the NLO contribution to the LO contribution for different terms in the moment $\mathcal{M}^0$ (eq.~\eqref{moment}), shown as a function of $s_0$ for various $\tau ({\small\text{GeV}^{-2}})$. (a) The NLO perturbative contribution is comparable to the LO perturbative contribution. (b) The NLO $\langle GG\rangle$ contribution is relatively small compared to the LO $\langle GG\rangle$ contribution. The corresponding ratio for the $\langle \bar{q}q\rangle^2$ contribution is not shown, as it does not contribute to the imaginary part of the correlator at leading order. \label{nlo_lo_ratio}}
\end{figure}

Overall, the NLO correction significantly lowers the predicted mass by $\sim0.7\,\text{GeV}$. This dramatic shift is due to the slow convergence of the perturbative series, where the NLO corrections are substantial, amounting to roughly 70\% of the LO contribution, as shown in figure~\ref{nlo_lo_ratio}.

\subsection{Impact of Vector Meson}

To quantify the impact of the $\rho$ or $\phi$ meson on the hybrid mass estimation, the coupling of the vector meson to the hybrid current $J^\mu$ must be estimated. Consider the dispersion relation for off-diagonal correlator~\eqref{rho-Hv_correlator}:
\begin{equation}
	\frac{1}{\pi}\text{Im}\Pi_v^{V H} = f_V f_1 \delta(s-m_v^2)+\theta(s-s_0)\rho(s),
\end{equation}
where $f_V$ is the vector meson coupling constant defined as
\begin{equation}
	\langle 0|J_V^\mu|V(q)\rangle = f_V \epsilon^\mu(q),
\end{equation}
while $f_1$ is the coupling of the vector meson to the hybrid current, defined in eq.~\eqref{H_coupling}. This can be estimated based on eq.~\eqref{rho-Hv_correlator} after applying the Borel transformation:
\begin{equation}
	f_1=\frac{e^{m_v^2\tau}}{\pi f_V}\int_0^{s_0} ds\,e^{-s\tau}\,\text{Im}\Pi_v^{VH}(s)
\end{equation}

For the $\rho$ meson, we take $m_\rho\simeq0.775\text{GeV}$ and $f_V=f_\rho m_\rho\simeq0.171\text{GeV}^2$~\cite{qsr,pdg,amp_decay}; while for the $\phi$ meson, we take $m_\phi\simeq1.019\text{GeV}$ and $f_V=f_\phi m_\phi\simeq0.235\text{GeV}^2$~\cite{pdg,WZ_decay}. To estimate $f_1$, according to refs.~\cite{qsr,rho_1--}, we chose a conservative range of $\tau$, $0.8\text{GeV}^{-2}<\frac{1}{\tau}<2\text{GeV}^{-2}$, and a conservative range of $s_0$, $1\text{GeV}^2<s_0<2\text{GeV}^2$, for $u\bar{u}$ configuration~\cite{qsr,rho_1--,quark-resonance_model}; while for $s\bar{s}$ configuration, we chose $1.5\text{GeV}^2<s_0<2.5\text{GeV}^2$~\cite{qsr_rho_omega_phi}.

\begin{figure}[t!]
	\centering
	\includegraphics[width=6.8cm]{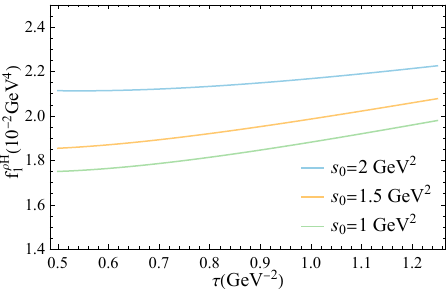}
	\hspace{\fill}
	\includegraphics[width=6.8cm]{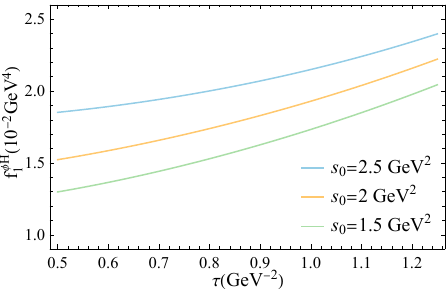}\\
	\vspace{-0.2cm}
	\caption{Estimated coupling strength $f_1$. Left: $\rho$ meson coupling to hybrid; Right: $\phi$ meson coupling to hybrid.}
	\label{f_V-H}
\end{figure}

As shown in figure~\ref{f_V-H}, we obtain $f_1\simeq0.0195\pm0.002 \text{GeV}^4$ for the $\rho$ coupling to the hybrid current. For the $\phi$ coupling to the hybrid, we find $f_1\simeq0.018\pm0.005\text{GeV}^4$. The former yields $\lambda^J_\rho=f_1/m_\rho^3\approx 37-46\,\text{MeV}$, which is comparable to the value of $\lambda^J_\rho$ in ref~\cite{rho_1--}.

\subsection{Gaussian Sum Rules}\label{GSR}

The Gaussian sum rule (GSR)~\cite{gaussian,gsr_bayesian} enables a more detailed examination of the spectral function. The GSR employs a Gaussian kernel, $e^{-(s-t)^2/4\tau}$, which focuses on a specific region of the spectrum. By varying the center of this kernel, one can, in principle, probe different regions of the spectrum and study excited states. Moreover, the GSR utilizes a fitting procedure to extract hadronic parameters, which reduces the bias from the choice of the continuum threshold $s_0$, and can yield a more robust prediction.

The GSR for the $1^{--}$ hybrid is defined as:
\begin{subequations}
	\begin{equation}
		G(\tau,s,s_0) = \frac{1}{\sqrt{4\pi\tau}}\int_0^{s_0}dt\, e^{-\frac{(s-t)^2}{4\tau}}\frac{1}{\pi}\text{Im}\Pi_v(t).
	\end{equation}
	On the hadronic side (two-resonance spectrum), applying the Gaussian transformation yields:
	\begin{equation}
		G_h(\tau,s)=\frac{1}{\sqrt{4\pi\tau}}\Big(f_1^2e^{-\frac{(s-m_1^2)^2}{4\tau}}+f_2^2 e^{-\frac{(s-m_2^2)^2}{4\tau}}\Big).
		\label{gsr_pole_s}
	\end{equation}
	\label{gsr_unnorm}
\end{subequations}
The renormalization-group improved GSR is obtained by setting $\mu^2=\sqrt{\tau}$~\cite{gaussian}. It is necessary to keep the width (resolution) of the Gaussian kernel finite. A very sharp Gaussian ($\tau\rightarrow0$) would require knowing the spectral function exactly, which is beyond the current theoretical capacity. A sufficiently wide Gaussian kernel also ensures that a resonance in the spectrum can be approximated by a $\delta$-function. In this work, we choose $\tau=8\,\text{GeV}^4$. The predicted masses are not sensitive to the specific choice of $\tau$, e.g. if choosing $\tau=10\,\text{GeV}^4$ as in ref.~\cite{1--_GSR_chen}, the masses in table~\ref{GSR_mass} would increase by only $0.01\,\text{GeV}$ for the NLO scenario.

The mass of hybrid $m_2$ and coupling strength $f_2$ can then be estimated by minimizing the $\chi^2$:
\begin{equation}
	\chi^2(\tau,s_0)=\sum_{i=0}^N  \big[G(\tau,s_i,s_0) - G_h(\tau,s_i)\big]^2.
	\label{chi_square}
\end{equation}
Here,
\begin{equation}
	N=\frac{s_\text{max} - s_\text{min}}{\delta s}+1,\quad s_i=s_\text{min}+i\times \delta s,
\end{equation}
and we choose
\begin{equation*}
	s_\text{min}=-15\,\text{GeV}^2,\quad s_\text{max}=30\,\text{GeV}^2,\quad \delta s = 0.2\,\text{GeV}^2.
\end{equation*}
The values of $s_\text{min}$ and $s_\text{max}$ are chosen to be far from the peak of $G(\tau,s,s_0)$, while $\delta s$ is chosen to be small enough to capture the detailed structure of $G(\tau,s,s_0)$.

\begin{figure*}[t!]
	\raggedright
	\hspace*{-0.25cm}
	\begin{subfigure}{0.48\textwidth}
		\includegraphics[width=6.95cm, height=4.1cm]{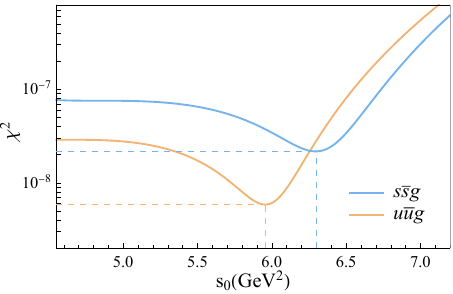}
		\caption{Minimized $\chi^2$ versus $s_0$.}
		\label{err_d8}
	\end{subfigure}
	\hspace*{\fill}	
	\begin{subfigure}{0.48\textwidth}
		\includegraphics[width=6.8cm, height=4.1cm]{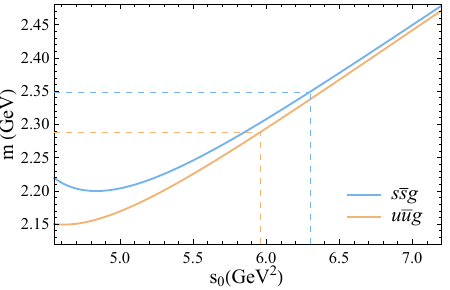}
		\caption{Fitted mass $m_2$ versus $s_0$.}
		\label{m_d8}
	\end{subfigure}\\[0.3cm]
	\begin{subfigure}{0.48\textwidth}
		\includegraphics[width=6.8cm, height=4.1cm,trim={0 0cm 0 0}]{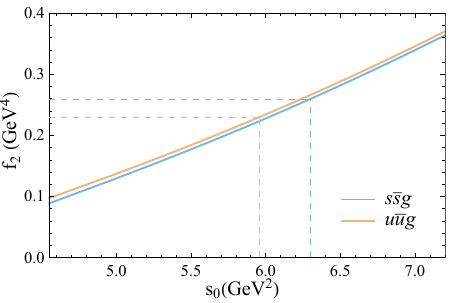}
		\caption{Fitted coupling strength $f_2$ versus $s_0$.}
		\label{fit_ss_d8}
	\end{subfigure}
	\hspace*{\fill}
	\begin{subfigure}{0.48\textwidth}
		\includegraphics[width=6.8cm, height=4.1cm,trim={0 0cm 0 0}]{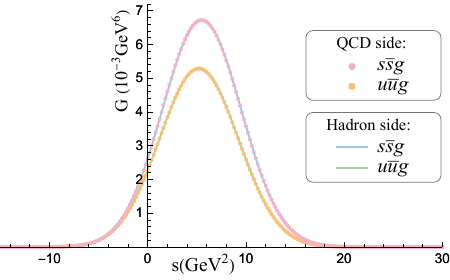}
		\caption{Optimal fitting for $u\bar{u}g$ and $s\bar{s}g$.}
		\label{fit_uu_d8}
	\end{subfigure}
	\caption{Next-to-leading order GSR fitting results, obtained by fitting eq.~\eqref{gsr_unnorm}. (a), (b), (c): Dashed lines mark the optimal value of $s_0$, as well as the corresponding $\chi^2$, predicted mass, and coupling $f_2$. (d): The optimal fitting of $G_h(\tau,s)$ (curves) to $G(\tau,s,s_0)$ (dots).\label{GSR_d8}}%
\end{figure*}


To visualize the solution's dependence on the continuum threshold $s_0$, we vary $s_0$ and, for each value, minimize the $\chi^2$ with respect to the $m_2$ and $f_2$. As shown in figure~\ref{GSR_d8}, the predicted masses $\sim 2.3\text{GeV}$ agree well with the LSR results in figure.~\ref{rho-ratio_k11}; the coupling $f_2\sim 0.25\text{GeV}$ then yields $f_1^2/f_2^2\sim 1/200$, indicating that the vector meson's contribution is nearly negligible.

To estimate the error of hybrid mass prediction, a Monte Carlo uncertainty analysis is performed as detailed in Appendix~\ref{monte_carlo}. However, we find that the fitting procedure is sensitive to the input QCD parameters and prone to failure. One can make the mass estimation more robust by normalizing eq.~\eqref{gsr_unnorm} as in ref~\cite{1--_GSR_chen}. Nevertheless, it is difficult to obtain a reliable coupling strength $f_2$ from the nearly negligible relative couplings $r=f_1^2/(f_1^2+f_2^2)$. Since our goal is to predict the mass of hybrid, we use normalized single resonance Gaussian sum rules instead~\cite{gaussian_norm,1--_GSR_chen}:
\begin{equation}
	\begin{split}
		\widetilde{G}(\tau,s,s_0)=&\,\,\frac{G(\tau,s,s_0)}{\int_{-\infty}^\infty ds\,G(\tau,s,s_0)},\\ \widetilde{G}_h(\tau,s)\,\,\,\,=&\,\,\frac{G_h(\tau,s)}{\int_{-\infty}^\infty ds\,G_h(\tau,s)} = \frac{1}{\sqrt{4\pi\tau}}e^{-\frac{(s-m^2)^2}{4\tau}}.
	\end{split}	
	\tag{3.8'}
	\label{gsr_norm}
\end{equation}
Here, the involved parameters are much less than in eq.~\eqref{gsr_unnorm}, making the mass prediction more robust. As summarized in table~\ref{GSR_mass}, the GSR analysis yields $1^{--}$ hybrid masses $\simeq2.3\,\text{GeV}$, which are consistent with the results in figures.~\ref{rho-ratio_k11} and \ref{GSR_d8}, and comparable to the experimentally measured mass of $\phi(2170)$. On the other hand, the GSR results also show that the mass of $u\bar{u}g$ is lower than the mass of $s\bar{s}g$ by $\simeq0.06\,\text{GeV}$, due to the smaller mass of $u$-quark.

\begin{table}[t!]
	\centering
	\caption{Single-resonance GSR fitting results in different scenarios. The values shown are determined via a Monte Carlo uncertainty analysis (the $\langle \bar{q}q\rangle$ and $\langle \bar{s}s\rangle$ are assumed to have a 10\% uncertainties). The $\overline{\chi_\text{min}^2}$ denotes the average minimum $\chi^2$.\label{GSR_mass}}
	\renewcommand{\arraystretch}{1.2}
	\begin{NiceTabular}{w{c}{1.8cm}w{c}{2cm}w{c}{2cm}w{c}{2cm}w{c}{2cm}}[hvlines]
		\noalign{\hrule height 1pt}
		\Block{2-1}{{\small Parameters}}&\multicolumn{2}{c|}{NLO}&\multicolumn{2}{c|}{LO}\\
		&$s\bar{s}g$&$u\bar{u}g$&$s\bar{s}g$&$u\bar{u}g$\\
		$\overline{\chi_\text{min}^2}$&$1.17\times10^{-5}$&$2.32\times10^{-5}$&$5.99\times10^{-5}$&$4.83\times10^{-5}$\\
		$s_0(\text{GeV}^2)$&$6.15\pm1.21$&$5.85\pm1.21$&$8.91\pm1.43$&$8.45\pm1.52$\\
		$m(\text{GeV})$&$2.31\pm0.23$&$2.25\pm0.23$&$2.77\pm0.22$&$2.70\pm0.24$\\
		\noalign{\hrule height 1pt}
	\end{NiceTabular}
\end{table}


\subsection{Robustness of the Results}\label{consist}

\begin{figure}[t!]
	\raggedright
	\hspace*{-0.5cm}
	\begin{subfigure}{0.43\textwidth}
		\includegraphics[width=7.8cm]{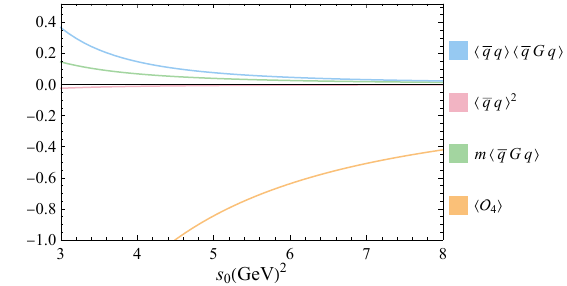}
		\caption{$s\bar{s}g$.}
	\end{subfigure}
	\hspace*{1.05cm}	
	\begin{subfigure}{0.43\textwidth}
		\includegraphics[width=7.8cm]{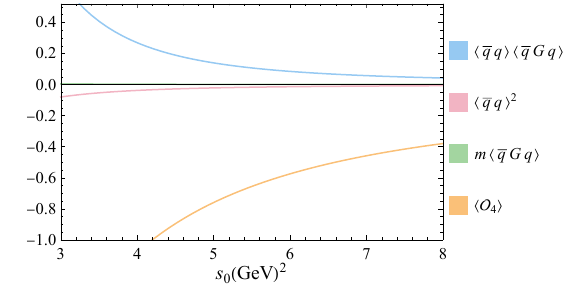}
		\caption{$u\bar{u}g$.}
	\end{subfigure}\\[-0.1cm]
	\caption{Ratios of each condensate's contribution to the perturbative contribution in the moment $\mathcal{M}^0$ (eq.~\eqref{moment}), as a function of $s_0({\small\text{GeV}^2})$. Here, NLO corrections are included, and $\tau=0.3{\small\text{GeV}^{-2}}$ is used. $\langle \mathcal{O}_4\rangle$ denotes the combined contribution of $m\langle \bar{q}q\rangle$ and $\langle GG\rangle$. The overall OPE converges well in each case.\label{cond_per_ratio}}
\end{figure}

\begin{figure}[t!]
	\raggedright
	\hspace*{-0.2cm}
	\begin{subfigure}{0.43\textwidth}
		\includegraphics[width=6.3cm]{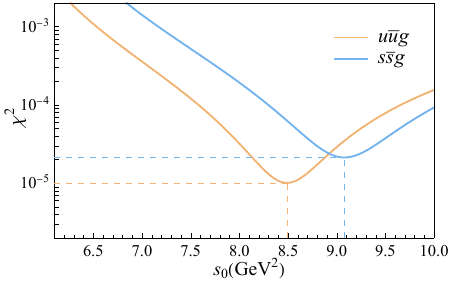}
		\vspace*{-0.1cm}
		\caption{Minimized $\chi^2$ versus $s_0$.}
		\label{err_lo}
	\end{subfigure}
	\hspace*{1.5cm}	
	\begin{subfigure}{0.43\textwidth}
		\includegraphics[width=6.3cm]{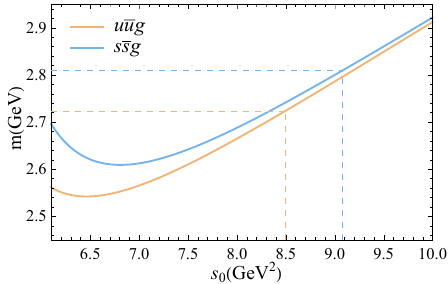}
		\vspace*{-0.1cm}
		\caption{Fitted mass $m$ versus $s_0$.}
		\label{m_lo}
	\end{subfigure}\\ 
	\caption{LO single-resonance GSR results, follow the same convention as in figure~\ref{GSR_d8}.\label{GSR_lo}}
\end{figure}

To assess the reliability of our mass determinations, we perform several cross-checks. As shown in figure~\ref{cond_per_ratio}, the OPE converges well, which implies that the corrections from condensates not included are small. For completeness, the LO GSR analysis is presented in figure~\ref{GSR_lo} and table~\ref{GSR_mass}. In both the LO and NLO analyses, the masses and continuum thresholds $s_0$ extracted from the LSR and GSR agree within uncertainties.

For dimension-6 and -8 condensates, the factorization (vacuum-saturation hypothesis) introduces large uncertainties. To assess the impact of deviations from this factorization, we introduce deviation factors $\kappa_6$ and $\kappa_8$ into the condensates, i.e. rewriting $\langle\bar{q}q\rangle^2$ and $\langle\bar{q}q\rangle\langle\bar{q}Gq\rangle$ as $\kappa_6\langle\bar{q}q\rangle^2$ and $\kappa_8 \langle\bar{q}q\rangle\langle\bar{q}Gq\rangle$, respectively. The precise values of $\kappa_6$ and $\kappa_8$ are not known, and they depend on the specific Dirac structures~\cite{Chetyrkin_1988,PhysRevD.91.074008}. For the scenarios considered in Refs~\cite{deviation_factors,deviation_factor_d6}, the author suggests that $\kappa_6\simeq3$ and $2\leq\kappa_8\leq5$. To see how the results change with large $\kappa_6$ and $\kappa_8$, we therefore examine the cases of $\kappa_6=3$ with $\kappa_8=2$ and $\kappa_8=5$. The corresponding results are summarized in table~\ref{GSR_mass_kappa}.

\begin{figure}[p!]
	\raggedright
	\includegraphics[width=7.05cm,height=3.8cm]{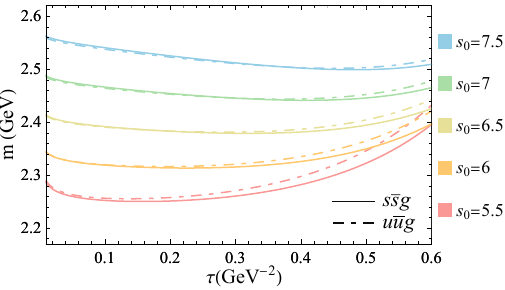}
	\hspace*{\fill}
	\includegraphics[width=7.05cm,height=3.8cm]{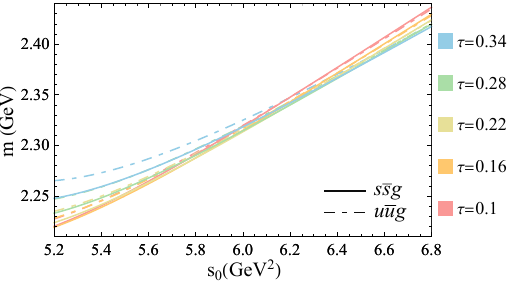}
	\vspace{-0.1cm}
	\caption{NLO LSR results using eq.~\eqref{ratio_rho} when $\kappa_6=3$ and $\kappa_8=2$: ${\small\sqrt{\mathcal{R}^0}}$ versus $\tau({\small\text{GeV}^{-2}})$ and ${\small s_0(\text{GeV}^2)}$.}
	\label{LSR_r0_k32}
\end{figure}

\begin{figure}[p!]
	\raggedright
	\hspace*{-0.25cm}
	\begin{subfigure}{0.43\textwidth}
		\includegraphics[width=6.25cm,height=3.8cm]{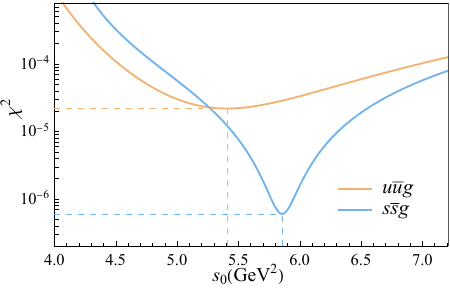}
		\vspace*{-0.1cm}
		\caption{Minimized $\chi^2$ versus $s_0$.}
	\end{subfigure}
	\hspace*{1.5cm}	
	\begin{subfigure}{0.43\textwidth}
		\includegraphics[width=6.22cm,height=3.8cm]{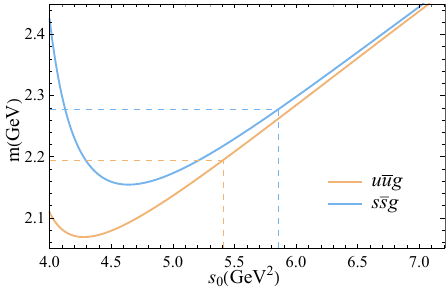}
		\vspace*{-0.1cm}
		\caption{Fitted mass $m$ versus $s_0$.}
	\end{subfigure}\\[-0.1cm]
	\caption{NLO single-resonance GSR fitting results when $\kappa_6=3$ and $\kappa_8=2$,  following the same convention as in figure~\ref{GSR_d8}.\label{GSR_nlo_k32}}
\end{figure}

Compared to the $\kappa_6=\kappa_8=1$ case, for $\kappa_6=3$ and $\kappa_8=2$, the LSR mass predictions are slightly higher (figures~\ref{LSR_r0_d8} and \ref{LSR_r0_k32}). Although the $u\bar{u}g$ appears incorrectly heavier than $s\bar{s}g$ at large $\tau$, the stable region yields a mass prediction of $\sim2.3\,\text{GeV}$ for both $u\bar{u}g$ and $s\bar{s}g$, which is still in agreement with the GSR results (figure~\ref{GSR_nlo_k32} and table~\ref{GSR_mass_kappa}).

\begin{figure}[p!]
	\raggedright
	\includegraphics[width=7.05cm,height=3.8cm]{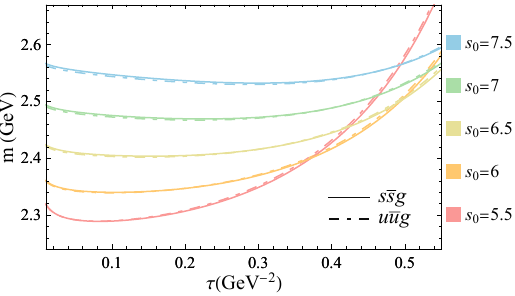}
	\hspace*{\fill}
	\includegraphics[width=7.05cm,height=3.8cm]{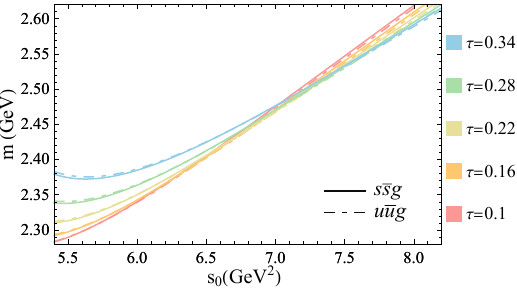}
	\caption{NLO LSR results using eq.~\eqref{ratio_rho} when $\kappa_6=3$ and $\kappa_8=5$: ${\small\sqrt{\mathcal{R}^0}}$ versus $\tau({\small\text{GeV}^{-2}})$ and ${\small s_0(\text{GeV}^2)}$.}
	\label{LSR_r0_k35}
\end{figure}

\begin{figure}[p!]
	\raggedright
	\hspace*{-0.35cm}
	\begin{subfigure}{0.43\textwidth}
		\includegraphics[width=6.25cm,height=3.8cm]{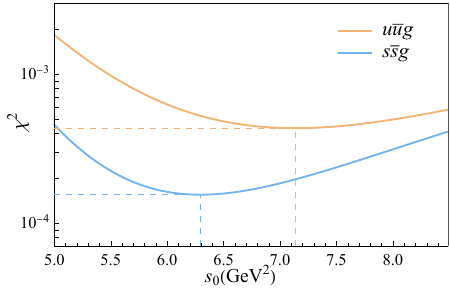}
		\caption{Minimized $\chi^2$ versus $s_0$.}
	\end{subfigure}
	\hspace*{1.5cm}	
	\begin{subfigure}{0.43\textwidth}
		\includegraphics[width=6.2cm,height=3.8cm]{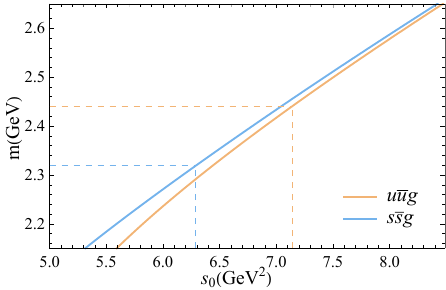}
		\caption{Fitted mass $m$ versus $s_0$.}
	\end{subfigure}\\[0.1cm]
	\caption{NLO single-resonance GSR fitting results when $\kappa_6=3$ and $\kappa_8=5$,  following the same convention as in figure~\ref{GSR_d8}.\label{GSR_nlo_k35}}
\end{figure}

For $\kappa_6=3$ and $\kappa_8=5$, the LSR predicts heavier masses $\sim2.4\text{GeV}$ (figure~\ref{LSR_r0_k35}), which agree with the GSR results shown in table.~\ref{GSR_mass_kappa}. However, the GSR incorrectly predicts $u\bar{u}g$ to be heavier than $s\bar{s}g$. This inconsistency arises because such a large $\kappa_8$ disrupts OPE convergence. We therefore discard this scenario. Even in this case, the mass predictions remain $\sim0.6\,\text{GeV}$ lower than the LO values and are broadly consistent with lattice QCD calculations~\cite{1--_lattice_dudek,1--_lattice_liu}. These analyses indicate that the contributions of dimension-6 and -8 four-quark condensates are small in the correlators, thereby ensuring the robustness of these mass predictions.


\begin{table}[t!]
	\centering
	\caption{Single-resonance NLO GSR fitting results for different factorization deviation factors. Obtained in a same way as in table~\ref{GSR_mass}.\label{GSR_mass_kappa}}
	\renewcommand{\arraystretch}{1.2}
	\begin{NiceTabular}{w{c}{1.8cm}w{c}{2cm}w{c}{2cm}w{c}{2cm}w{c}{2cm}}[hvlines]
		\noalign{\hrule height 1pt}
		\Block{2-1}{{\small Parameters}}&\multicolumn{2}{c|}{$\kappa_6=3$, $\kappa_8=2$}&\multicolumn{2}{c|}{$\kappa_6=3$, $\kappa_8=5$}\\
		&$s\bar{s}g$&$u\bar{u}g$&$s\bar{s}g$&$u\bar{u}g$\\
		$\overline{\chi_\text{min}^2}$&$2.85\times10^{-5}$&$7.28\times10^{-5}$&$1.74\times 10^{-4}$&$4.30\times 10^{-4}$\\
		$s_0(\text{GeV}^2)$&$6.03\pm1.21$&$6.02\pm1.31$&$6.40\pm1.60$&$7.03\pm2.16$\\
		$m(\text{GeV})$&$2.29\pm0.23$&$2.28\pm0.24$&$2.34\pm0.28$&$2.43\pm0.37$\\
		\noalign{\hrule height 1pt}
	\end{NiceTabular}
\end{table}

\section{Conclusion}\label{conclusion}

We present a comprehensive next-to-leading-order (NLO) QCD sum rule analysis for the light vector hybrid meson, revealing that NLO corrections are substantial and essential for an accurate mass prediction. These corrections systematically lower the predicted mass, resolving a long-standing tension between different theoretical approaches.

Our results, $2.31\pm0.23$ GeV for $s\bar{s}g$ and $2.25\pm0.23$ GeV for $u\bar{u}g$, reconcile QCD sum rule predictions with both lattice QCD ($2.2-2.5$ GeV~\cite{1--_lattice_dudek,1--_lattice_liu}) and the flux-tube model ($1.8-2.2$ GeV~\cite{1--_flux_tube}). Crucially, our prediction for the strangeonium hybrid mass is in agreement with the experimental data for the $\phi(2170)$ resonance. This agreement provides compelling evidence that the $\phi(2170)$ is a prime candidate for the elusive light vector hybrid meson, or is a mixed state with a large hybrid component.

\appendix


\section{Gluon Propagators and Cancellation of IR Pole\label{gluon_IR_pole}}
Following a procedure similar to that in ref.~\cite{high_order_condensates}, the gluon propagator in the background gluon field can be derived as follows. Replace gluon field with $a_\mu^n + A_\mu^n$, where $a_\mu$ is the quantum gluon field and $A_\mu$ is the background gluon field. By choosing the gauge-fixing term as $-\frac{1}{2}(\partial_\mu a_\nu)^2$, the required gluon vertices are~\cite{1-+_jin}:
\begin{equation}
	\begin{split}
		\begin{tikzpicture}[scale=1.5, baseline=-\the\dimexpr\fontdimen22\textfont2\relax]
			\begin{feynman}
				\vertex (a0);
				\vertex [left=0.1cm of a0](al);
				\vertex [right=1cm of a0](c);
				\vertex [above=0.7cm of c](cu);
				\vertex [right=0.8cm of a0](cl);
				\vertex [right=1.2cm of a0](cr);
				\vertex [above=0.7cm of cl](clu);
				\vertex [above=0.7cm of cr](cru);
				\vertex [right=2.0cm of a0](b0);
				\vertex [right=0.1cm of b0](br);
				\diagram*[small]{
					(b0)--[photon](c)--[photon](a0),
					(cu)--[photon, insertion={[size=1.5pt]0}](c)
				};
				\draw(al) node[scale=0.7, transform shape]{$^a_\alpha$};
				\draw(br) node[scale=0.7, transform shape]{$\,\,^b_\beta$};
				\draw(0.65,-0.2) node[scale=0.55, transform shape]{$x$};
				\draw(1.07,0.11) node[scale=0.55, transform shape]{$\leftarrow q$};
			\end{feynman}
		\end{tikzpicture} &= iG_{\alpha\beta} + (q^\rho A_\rho + A_\rho q^\rho)g_{\alpha\beta}-(q_\beta A_\alpha + A_\beta q_\alpha),\\
		\begin{tikzpicture}[scale=1.5, baseline=-\the\dimexpr\fontdimen22\textfont2\relax]
			\begin{feynman}
				\vertex (a0);
				\vertex [left=0.1cm of a0](al);
				\vertex [right=1cm of a0](c);
				\vertex [above=0.7cm of c](cu);
				\vertex [right=0.7cm of a0](cl);
				\vertex [right=1.3cm of a0](cr);
				\vertex [above=0.7cm of cl](clu);
				\vertex [above=0.7cm of cr](cru);
				\vertex [right=2.0cm of a0](b0);
				\vertex [right=0.1cm of b0](br);
				\diagram*[small]{
					(b0)--[photon](c)--[photon](a0),
					(clu)--[photon, insertion={[size=1.5pt]0}](c),
					(cru)--[photon, insertion={[size=1.5pt]0}](c)
				};
				\draw(al) node[scale=0.7, transform shape]{$^a_\alpha$};
				\draw(br) node[scale=0.7, transform shape]{$\,\,^b_\beta$};
				\draw(0.65,-0.2) node[scale=0.55, transform shape]{$x$};
				\draw(1.07,0.11) node[scale=0.55, transform shape]{$\leftarrow q$};
			\end{feynman}
		\end{tikzpicture} &= i g^2 f^{anc}f^{cmb}A^{n\,\rho}A^m_\rho g_{\alpha\beta} -ig^2 f^{anc}f^{cmb}A^m_\alpha A^n_\beta,
	\end{split}
	\label{gluon_v_partial_d}
\end{equation}
where we adopt the matrix notations $G_{\mu\nu} = g f^{anb} G^n_{\mu\nu}(x)$ and $A_\mu=g f^{anb} A^n_\mu(x)$. The ordering of background gluon field $A$ and momentum $q$ is important, since
\begin{equation}
	A^n_\mu(x)=\frac{1}{2} x^\rho G^n_{\rho\mu}(0)+\frac{1}{3\cdot 1!} x^\rho x^\eta D_\rho G^n_{\eta\mu}(0)\cdots
\end{equation}
in Fock-Schwinger gauge~\cite{high_order_condensates}, and $x$ can be written as $-i\partial_q$ or $i\overleftarrow{\partial_q}$ in the propagator.

Alternatively, choosing the commonly adopted gauge fixing term (background gauge) $-\frac{1}{2}(D_\mu a_\nu)^2$ in QCD sum rules~\cite{high_order_condensates}, where $D_\mu = \partial_\mu +g f^{anb}A_\mu^n$, the gluon vertices become:
\begin{equation}
	\begin{split}
		\begin{tikzpicture}[scale=1.5, baseline=-\the\dimexpr\fontdimen22\textfont2\relax]
			\begin{feynman}
				\vertex (a0);
				\vertex [left=0.1cm of a0](al);
				\vertex [right=1cm of a0](c);
				\vertex [above=0.7cm of c](cu);
				\vertex [right=0.8cm of a0](cl);
				\vertex [right=1.2cm of a0](cr);
				\vertex [above=0.7cm of cl](clu);
				\vertex [above=0.7cm of cr](cru);
				\vertex [right=2.0cm of a0](b0);
				\vertex [right=0.1cm of b0](br);
				\diagram*[small]{
					(b0)--[photon](c)--[photon](a0),
					(cu)--[photon, insertion={[size=1.5pt]0}](c)
				};
				\draw(al) node[scale=0.7, transform shape]{$^a_\alpha$};
				\draw(br) node[scale=0.7, transform shape]{$\,\,^b_\beta$};
				\draw(0.65,-0.2) node[scale=0.55, transform shape]{$x$};
				\draw(1.07,0.11) node[scale=0.55, transform shape]{$\leftarrow q$};
			\end{feynman}
		\end{tikzpicture} &= 2iG_{\alpha\beta} + (q^\rho A_\rho + A_\rho q^\rho)g_{\alpha\beta}\\
		\begin{tikzpicture}[scale=1.5, baseline=-\the\dimexpr\fontdimen22\textfont2\relax]
			\begin{feynman}
				\vertex (a0);
				\vertex [left=0.1cm of a0](al);
				\vertex [right=1cm of a0](c);
				\vertex [right=0.7cm of a0](cl);
				\vertex [right=1.3cm of a0](cr);
				\vertex [above=0.7cm of cl](clu);
				\vertex [above=0.7cm of cr](cru);
				\vertex [right=2.0cm of a0](b0);
				\vertex [right=0.1cm of b0](br);
				\diagram*[small]{
					(b0)--[photon](c)--[photon](a0),
					(clu)--[photon, insertion={[size=1.5pt]0}](c),
					(cru)--[photon, insertion={[size=1.5pt]0}](c)
				};
				\draw(al) node[scale=0.7, transform shape]{$^a_\alpha$};
				\draw(br) node[scale=0.7, transform shape]{$\,\,^b_\beta$};
				\draw(0.65,-0.2) node[scale=0.55, transform shape]{$x$};
				\draw(1.07,0.11) node[scale=0.55, transform shape]{$\leftarrow q$};
			\end{feynman}
		\end{tikzpicture} &= i g^2 f^{anc}f^{cmb}A^{n\,\rho}A^m_\rho g_{\alpha\beta}.
	\end{split}
	\label{gluon_v_covariant_d}
\end{equation}

Eq.~\eqref{gluon_v_covariant_d} is more convenient for calculations, whereas eq.~\eqref{gluon_v_partial_d} makes the cancellation of the $\log/\varepsilon$ pole in figure~\ref{c_gg} more obvious because both figure~\ref{c_gg} and eq.~\eqref{ren_hybrid} are evaluated in Feynman gauge. Each diagram in figure~\ref{hybrid_ren} corresponds to a subdiagram in figure~\ref{c_gg} that contributes to the $\log/\varepsilon$ pole, with the exception of the diagrams
\begin{equation}
	\includegraphics[width=5cm, valign=c]{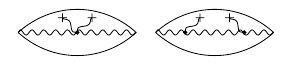}\quad\text{and}\quad\includegraphics[width=2.1cm, valign=c]{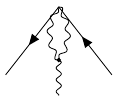}
	\label{IR_sub}
\end{equation}
in figure~\ref{c_gg} and \ref{hybrid_ren}, respectively. To ensure cancellation of the $\log/\varepsilon$ pole, these diagrams must be related. The IR pole in the former is canceled by the diagrams corresponding to the counterterm: 
\begin{equation}
	(C_0 + Z_3^{-\frac{1}{2}}Z_g^{-1}-1)\,\overline{\Psi}_{f_a} \,g_0 \Gamma G^{\mu\nu} \Psi_{f_b},
	\label{IR_cancel}
\end{equation}
where $C_0$ originates from the latter in eq.~\eqref{IR_sub}. The presence of $Z_3$ and $Z_g$ is justified, because no other $\langle GG\rangle$ diagram in figure~\ref{c_gg} relates to the $Z_3$ in eq.~\eqref{ren_hybrid}, but $Z_3$ introduces a $n_f$-dependency. The combination of $Z_3^{-\frac{1}{2}}Z_g^{-1}$ then ensures the result is independent of $n_f$, as the diagrams in figures~\ref{IR_sub} and~2 are $n_f$-independent.


\section{Evaluation of Dimension-8 Four-Quark Condensates\label{d8_factor}}

For simplicity, we adopt the following notations:
\begin{equation}
	D_\rho G_{\alpha\nu}=gT^n D_\rho^{nm}G^{m}_{\alpha\nu},\qquad G_{\alpha\nu}=gT^n G^n_{\alpha\nu},
\end{equation}
for the dimension-8 condensate. The last diagram in figure~\ref{c_d8} contributes to non-factorizable dimension-8 condensates, such as $i\langle \bar{q} D_\rho {G_\alpha}^\mu G_{\mu\nu} \gamma^\rho.\gamma^\alpha.\gamma^\nu q\rangle$ and $i\langle \bar{q} G_{\mu\nu}D_\rho G^{\mu\nu}\gamma^\rho q\rangle$, which can be rewritten using the Bianchi identity and the equation of motion~\cite{1-+_huang}. For example:
\begin{equation}
	\begin{split}
		\langle \bar{q} G^{\mu\alpha}D_\rho G_{\alpha\mu}\gamma^\rho q\rangle&=\langle \bar{q} {G_\mu}^\alpha D_\rho G_{\alpha\nu}\frac{1}{2}(\gamma^\rho.\gamma^\mu.\gamma^\nu+\gamma^\rho.\gamma^\nu.\gamma^\mu) q\rangle\\
		&=\frac{1}{2}\langle \bar{q} {G_\mu}^\alpha D_\rho G_{\alpha\nu}(2g^{\rho\mu}\gamma^\nu-\gamma^\mu.\gamma^\rho.\gamma^\nu+\gamma^\rho.\gamma^\nu.\gamma^\mu) q\rangle\\
		&=\langle \bar{q} D^\rho G_{\alpha\rho}G_{\alpha\nu}\gamma^\nu q\rangle +\langle \bar{q}\nabla^\rho (G_{\alpha\rho} G_{\alpha\nu})\gamma^\nu q\rangle\\
		&\qquad+\frac{1}{2}\langle \bar{q}{G_\mu}^\alpha D_\rho G_{\alpha\nu}(-\gamma^\mu.\gamma^\rho.\gamma^\nu+\gamma^\rho.\gamma^\nu.\gamma^\mu) q\rangle.
	\end{split}
\end{equation}
Applying the Bianchi identity, the last line becomes:
\begin{equation}
	\begin{split}
		&\frac{1}{4}\langle \bar{q}{G_\mu}^\alpha \big[(D_\rho G_{\alpha\nu}+D_\nu G_{\alpha\rho})-D_\alpha G_{\nu\rho}\big](-\gamma^\mu.\gamma^\rho.\gamma^\nu+\gamma^\rho.\gamma^\nu.\gamma^\mu) q\rangle\\
		&=\frac{1}{4}\langle \bar{q}{G_\mu}^\alpha (D_\rho G_{\alpha\nu}+D_\nu G_{\alpha\rho})(-\gamma^\mu g^{\rho\nu}+g^{\rho\nu}\gamma^\mu)q\rangle+\frac{1}{4}\langle \bar{q}{G_\mu}^\alpha  D_\alpha G_{\nu\rho}(\gamma^\mu.\gamma^\rho.\gamma^\nu-\gamma^\rho.\gamma^\nu.\gamma^\mu)q\rangle\\
		&=\frac{1}{4}\langle \bar{q} D^\alpha G_{\alpha\mu}  G_{\nu\rho}(\gamma^\mu.\gamma^\rho.\gamma^\nu-\gamma^\rho.\gamma^\nu.\gamma^\mu)q\rangle + \frac{1}{4}\langle \bar{q}\nabla^\alpha (G_{\alpha\mu}  G_{\nu\rho})(\gamma^\mu.\gamma^\rho.\gamma^\nu-\gamma^\rho.\gamma^\nu.\gamma^\mu)q\rangle,
	\end{split}
\end{equation}
where $\nabla_\mu=\partial_\mu-ig T^n A^n_\mu$ and $D^{nm}_\mu=\partial_\mu \delta^{nm}+g f^{nlm}A^l_\mu$. The first term yields a factorizable dimension-8 condensate, while the second vanishes if factorized into $m\langle \bar{q}q\rangle \langle GG\rangle$. After a tedious calculation, the last diagram in figure~\ref{c_d8} yields:
\begin{equation}
	\begin{split}
		&\Big(g^{\mu\nu}+\frac{q^\mu q^\nu}{2q^2}\Big)\Big(-\frac{i}{9q^2}\langle\bar{q}D^\alpha G_{\alpha\mu}G_{\nu\rho}\gamma^{\mu\nu\rho} q\rangle-\frac{4i}{9q^2}\langle\bar{q}D^\alpha G_{\alpha\rho}G_{\rho\mu}\gamma^\mu q\rangle\Big)	\\
		&=\Big(\frac{q^\mu q^\nu}{q^2}-g^{\mu\nu}\Big)\frac{25\pi}{81q^2}\alpha_s\langle \bar{q}q\rangle \langle \bar{q}Gq\rangle - \frac{q^\mu q^\nu}{q^4}\frac{25\pi}{54}\alpha_s\langle \bar{q}q\rangle \langle \bar{q}Gq\rangle,
	\end{split}
\end{equation}
where $\gamma^{\mu\nu\rho}=\gamma^{[\rho}.\gamma^\nu.\gamma^{\rho]}$.

The condensate in the second-to-last diagram in figure~\ref{c_d8} exhibits an ambiguity under factorization. Applying the equation of motion before factorization yields:
\begin{equation}
	\Big(\frac{q^\mu q^\nu}{q^2}-g^{\mu\nu}\Big)\frac{73\pi}{324q^2}\alpha_s\langle \bar{q}q\rangle \langle \bar{q}Gq\rangle + \frac{q^\mu q^\nu}{q^4}\frac{23\pi}{36}\alpha_s\langle \bar{q}q\rangle \langle \bar{q}Gq\rangle.
	\label{d8_em_vs}
\end{equation}
In contrast, applying the vacuum saturation hypothesis first gives:
\begin{equation}
	\Big(\frac{q^\mu q^\nu}{q^2}-g^{\mu\nu}\Big)\frac{2\pi}{9q^2}\alpha_s\langle \bar{q}q\rangle \langle \bar{q}Gq\rangle + \frac{q^\mu q^\nu}{q^4}\frac{2\pi}{3}\alpha_s\langle \bar{q}q\rangle \langle \bar{q}Gq\rangle.
	\label{d8_vs_em}
\end{equation}

The difference between these two results is small; for the vector part, the difference can be neglected. The results in eqs.~\eqref{Pi_v(s)} and~\eqref{Pi_s(s)} correspond to choosing eq.~\eqref{d8_em_vs}.

For eq.~\eqref{rho-Hv_correlator}, the $\langle \mathcal{O}_8\rangle$ in figure.~\ref{rho-H_diagrams} involves the expansion relation:
\begin{equation}
	\langle \bar{q}_i^a(0) G_{\alpha\beta}^n q_j^b(x)\rangle = \frac{(2-3d)C_A^2+6(d-2)}{96 (d-1) d (d+2) C_A^3C_F} \pi\alpha _s T^{n\,ba} x^2 (\gamma ^\beta x^\alpha-\gamma ^\alpha x^\beta)_{ji} \langle\bar{q}q\rangle  \langle\bar{q}Gq\rangle +\cdots,
	\label{d8_expansion}
\end{equation}
where the terms irrelevant to the dimension-8 condensate $\langle \bar{q}q\rangle\langle \bar{q}G q\rangle$ are omitted, and applying the equation of motion before or after factorization yields the same result.

\section{Monte Carlo Analysis for Error Estimation\label{monte_carlo}}

The numerical values in tables~\ref{GSR_mass} and \ref{GSR_mass_kappa} are obtained by Monte Carlo method. Based on the QCD parameters listed at the end of section.~\ref{NLO_correlator_1--}, we generate 2000 sets of input parameters by sampling from independent Gaussian distributions. The $\langle \bar{q}q\rangle$ and $\langle \bar{s}s\rangle$ are assigned a 10\% relative uncertainty. For each of the 2000 parameter sets, we perform the Gaussian Sum Rule fitting procedure, which yields 2000 optimal pairs of $(m,s_0)$. The total uncertainty ($\sigma$) is computed by combining two distinct sources of error in quadrature:
	\begin{equation}
		\sigma=\sqrt{\hat{\sigma}^2 + \overline{\Delta^2}}.
	\end{equation}
	Where:
	\begin{itemize}
		\item $\hat{\sigma}^2$ is the variance of the 2000 fitted values (for $m$ or $s_0$), representing the uncertainty propagated from the input QCD parameters.
		
		\item $\overline{\Delta^2}$ is the mean of the squared fitting errors from each individual fit. The error for a single fit, $\Delta$, is estimated from the Hessian matrix 
		\NiceMatrixOptions{cell-space-limits = 0.05cm}
		\begin{equation}
			H(\chi^2)=\begin{bNiceMatrix}
				\displaystyle\frac{\partial^2\chi^2}{\partial s^2_0} & \displaystyle\frac{\partial^2\chi^2}{\partial s_0\partial m}\\
				\displaystyle\frac{\partial^2\chi^2}{\partial s_0\partial m}&\displaystyle\frac{\partial^2\chi^2}{\partial m^2}\\
			\end{bNiceMatrix} 
		\end{equation}
		evaluated at the minimum of $\chi^2$, via
		\begin{equation}
			\Delta=\sqrt{2\chi^2_\text{min}\, H^{-1}(\chi^2_\text{min})}.
			\label{delta_err}
		\end{equation}
		This definition is independent of the grid size $\delta s$ in eq.~\eqref{chi_square}, provided that $\delta s$ is chosen sufficiently small. In the limit $\delta s\to0$, $\chi^2$ scales as $1/\delta s$ while $H^{-1}$ scales as $\delta s$. Eq.~\eqref{delta_err} quantifies the intrinsic uncertainty of the fitting method itself.
	\end{itemize}


\acknowledgments
This work is supported by the National Natural Science Foundation of China under Grant No. 12175318, 12305104, and 11175153, as well as the Education Department of Hunan Province under Grant No. 24B0503.

	\bibliographystyle{JHEP}
	\bibliography{refs}
\end{document}